\documentclass[structabstract]{aa} 
\usepackage{graphicx} 
\usepackage{float}
\usepackage{color}
\usepackage{txfonts}
\usepackage{array}
\usepackage{natbib}
\usepackage{units}
\usepackage{url}
\usepackage{lscape}
\usepackage{supertabular,booktabs}
\usepackage{comment}
\usepackage{multirow}
\usepackage{nccmath}
\usepackage{amsmath, amstext}
\usepackage[normalem]{ulem}

\begin{document}

\title{A revised distance to IRAS 16293$-$2422 from VLBA astrometry of associated water masers}
\author{S. A.\ Dzib\inst{1}
\and
G. N. Ortiz-Le\'on\inst{1,2}
\and
A. Hern\'andez-G\'omez\inst{3,4}
\and
L. Loinard\inst{3,5}
\and
A.~J.~Mioduszewski\inst{6}
\and
M.~Claussen\inst{6}
\and
K.~M.~Menten\inst{1}
\and
E.~Caux\inst{4}
\and 
A.~Sanna\inst{1}
}

\institute{Max-Planck-Institut f\"ur Radioastronomie, Auf dem H\"ugel 69,
 D-53121 Bonn, Germany 
 \and Humboldt Fellow
 \and Instituto de Radioastronom\'{\i}a y Astrof\'{\i}sica, Universidad Nacional Aut\'onoma de M\'exico, Morelia 58089, Mexico
 \and IRAP, Universit\'e de Toulouse, CNRS, UPS, CNES, Toulouse, France
 \and Instituto de  Astronom\'{\i}a, Universidad Nacional Aut\'onoma de M\'exico, Apartado Postal 70-264, CdMx C.P. 04510, Mexico \and
 National Radio Astronomy Observatory, P.O. Box 0, Socorro, NM 87801, USA
 \\
\email{sdzib, gortiz, kmenten, asanna @mpifr-bonn.mpg.de; a.hernandez, l.loinard  @irya.unam.mx; amiodusz, mclausse @nrao.edu; and
emmanuel.caux@irap.omp.eu
 }
}

\date{Received 2017; }
\abstract
{IRAS 16293-2422 is a very well studied young stellar system seen in projection towards the L1689N cloud in the 
Ophiuchus complex. However, its distance is still uncertain with a range of values from 120~pc to 180~pc.
Our goal is to measure the trigonometric parallax of this young star by means of H$_2$O maser emission.
We use archival data from 15 epochs of VLBA observations of the 22.2 GHz water maser line.
By modeling the displacement on the sky of the H$_2$O maser spots,  we derived a  trigonometric parallax 
of $7.1\pm0.6$ mas, corresponding to a distance of $141_{-21}^{+30}$ pc.
This new distance is in good 
agreement with recent values obtained for other magnetically active young stars in the L1689 cloud.
We relate the kinematics 
of these masers with the outflows and the recent ejections powered by source A in the system.
}

\keywords{astrometry --- masers --- stars:formation --- stars: individual (IRAS 16293-2422) 
--- techniques: interferometric}
\titlerunning{Astrometry to water masers in I16293}

\maketitle

\section{Introduction}\label{sec:intro}

\begin{figure*}[!ht]
   \centering
  \includegraphics[height=0.5\textwidth]{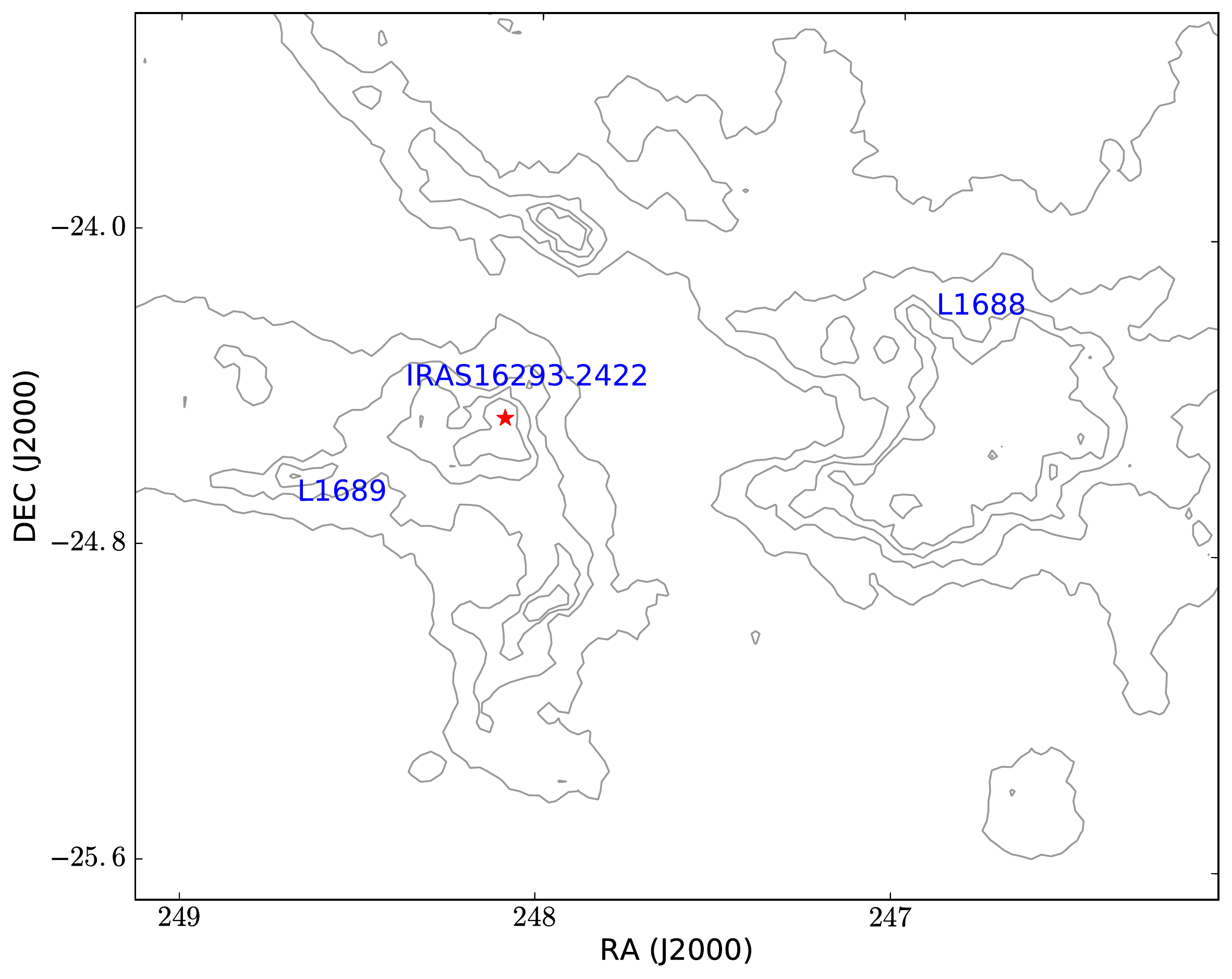}
   \caption{The Ophiuchus molecular cloud traced by optical extinction obtained as part of the COMPLETE project \citep{Ridge_2006}. The location of I16293, the core (L1688) and the Lynds 1689 (L1689) are indicated. The contours corresponds to $A_V=4,7,10$ and 13. }
   \label{fig:map}
\end{figure*}

As a result of the development and improvement of Very Long Base Interferometry (VLBI) techniques in recent decades,  it has become possible to measure the trigonometric parallax of deeply embedded young stellar objects (YSOs) with very high accuracy. This information is crucial for the determination of some of the  most fundamental stellar parameters, such as age, luminosity, and  mass. Low-mass YSOs with non-thermal emission (such as gyrosynchrotron or synchrotron emission) are good candidates to be observed with VLBI, since they have a high surface brightness over a few solar radii \citep[e.g.][]{loinard2008,dzib2016,OrtizLeon2017}. However, VLBI observations of Class 0 objects, the youngest stars, are more complicated because their non-thermal emission can suffer strong free-free absorption effects from the ionized winds that they typically power.  

Emission from water masers has been observed in a variety of objects and has been used to determine their trigonometric parallax and hence their distance  \citep[e.g.,][]{imai2007,hirota2008,reid2009,sanna2017Sci}. Although water masers can vary on short time scales \citep{claussen1996}, and their structure can be resolved at angular resolutions of VLBI instruments in nearby regions \citep{imai2007}, they represent a fundamental tool to determine distances to far away star-forming regions.  Indeed, recent VLBI observations of strong water masers have provided an accurate determination of the distance to a star forming region at 20 kpc from the Sun \citep{sanna2017Sci}.

Water maser emission has been observed towards IRAS~16293-2422 (I16293 hereafter), a very well studied young stellar system located to the north of the Lynds 1689 (L1689) cloud, in the  Ophiuchus complex (Figure \ref{fig:map}). This system is comprised by two main sources, A and B, identified from interferometric observations at centimeter wavelengths \citep{wootten1989,mundy1992} that are separated by 5$''$ and have properties of Class 0 objects, i.e., YSOs in very early evolutionary phases. Source A itself is resolved into two sub--sources, A1 and A2, separated by 0\rlap{.}$''$3 \citep{chandler2005,loinard2007}. Recently, other continuum sources associated with an episodic ejection from A were observed with the VLA \citep{loinard2007,pech2010,loinard2013}. Source A seems to power two outflows in the E-W and NW-SE directions observed through the emission of CO and SiO lines \citep{mizuno1990,girart2014}. There is also a compact outflow extending along the axis between A and B sources in the SE-NW direction. Since I16293 is an interesting laboratory to study the kinematics and dynamics of clustered star formation, it is of crucial importance to determine its distance very precisely and, therefore, infer correctly its physical parameters. Furthermore, relating the small scale kinematics of the water masers with those of the larger-scale outflows would be interesting.

Previously, \cite{imai2007} reported on the emission of H$_2$O masers observed with the VLBI Exploration of Radio Astrometry (VERA) array towards this source and obtained an annual parallax for a maser feature of 5.6$_{-0.5}^{+1.5}$ mas, corresponding to a distance of 178$_{-37}^{+18}$ pc. On the other hand, \cite{loinard2008} measured the distance to the core of Ophiuchus complex (also known as Lynds 1688 or L1688) from VLBA observations of non-thermal emission from two YSOs and found a distance of $120_{-4.2}^{+4.5}$ pc. This distance is now commonly used in the literature as the distance to I16293. More recently, \cite{OrtizLeon2017} measured the distance to the L1689 streamer using VLBA observations of three stellar systems, finding a mean parallax of 6.79$\pm$0.16 mas, which corresponds to a distance of 147.3$\pm$3.4 pc. They also revisited the mean distance to the Ophiuchus core and found a mean parallax of $7.28\pm0.06$ mas, corresponding to a distance of 137.3$\pm$1.2~pc. As the measured distance to I16293 measured by \cite{imai2007}  is significantly larger than that
to other stars in the L1689 cloud, it is not yet clear whether or not it is a background object seen along the line of sight to L1689. However,
it is hard to think that such a deeply embedded young star should not be part of a cloud when it appears in projection on the densest part of that cloud (see Fig.~\ref{fig:map}). It is clear that the distance to I16293 needs to be revisited. %

In this paper, we present high sensitivity VLBI observations of H$_2$O masers towards I16293. From these observations, we were able to identify the positions of several masers spots and perform a precise astrometry, finding a distance to I16293 that is in good agreement with the recent measured values to other young stars in the Ophiuchus complex.

\section{Observations and Data Calibration}

We analyze a series of 18 observations of water masers at 22.2~GHz  carried out with the Very Long Baseline Array (VLBA) as part of project BC152 (PI: M.~Claussen). These observations cover a period of eight months from 2005 August 2 to 2006 April 13 (see Table \ref{table1}), and were taken at intervals of about 15 days. 
Each epoch consisted of cycles switching between I16293 and  the quasar J1625-2527 ($\sim$1\rlap{.}$^\circ$8  away), with two minutes of integration time on the target and one minute on the quasar. Data were taken in right and left circular polarizations with four baseband channels (BBCs) of 8 MHz bandwidth each. The BBC containing the maser line, which  was centered at 22.2371~GHz, was also correlated with a channel separation of 15.6250~kHz (corresponding to a velocity resolution of 0.215 km~s$^{-1}$) to produce 512 spectral channels. Throughout the whole data analysis, we only considered  the BBC containing the maser emission. 

\begin{table*}[h!]
\centering
\small
\renewcommand{\arraystretch}{1.31}
\caption{Main parameters of all images obtained for each VLBA observation dataset. }
\label{table1}
\begin{tabular}{cclcc}
&&&&\\ \hline
Epoch & Date of  observation  & Synthesized beam     & Noise& Peak  \\ 
      &  (yyyy-mm-dd/hh:mm)   & (mas$\times$mas); PA & (mJy beam$^{-1}$) & (Jy beam$^{-1}$)  \\ \hline
A & 2005-Aug-02/03:00  &\multicolumn{1}{c}{...} & ... & ...\\
B & 2005-Aug-16/02:05  &\multicolumn{1}{c}{...} & ... & ...\\
C & 2005-Aug-30/01:10  &\multicolumn{1}{c}{...} & ... & ...\\
D & 2005-Sep-12/23:17  &1.14$\times$0.29; -20.1$^{\circ}$  & 19 & 0.35\\
E & 2005-Sep-25/22:19  &0.74$\times$0.25; -10.2$^{\circ}$  & 10 & 0.21\\
F & 2005-Oct-13/20:38  &0.77$\times$0.26; -10.9$^{\circ}$  & 40 & 0.55\\
G & 2005-Nov-01/19:55  &0.91$\times$0.29; -7.9$^{\circ}$   & 40 & 2.23 \\
H & 2005-Nov-12/19:12  &0.87$\times$0.27; -11.6$^{\circ}$  & 33 & 0.70 \\
I & 2005-Nov-27/18:14  &0.84$\times$0.25; -8.7$^{\circ}$   & 38 & 0.53\\
J & 2005-Dec-08/17:31  &1.33$\times$0.31; 7.2$^{\circ}$     & 21 & 1.10\\
K & 2005-Dec-22/16:34  &0.83$\times$0.26; -9.7$^{\circ}$   & 18 & 0.56\\
L & 2006-Jan-06/15:36  &0.78$\times$0.28; -8.9$^{\circ}$   & 15 & 0.69\\
M & 2006-Jan-18/14:53  &0.74$\times$0.25; -11.1$^{\circ}$   & 33 & 0.83\\
N & 2006-Feb-04/13:41  &0.81$\times$0.25; -11.0$^{\circ}$  & 28 & 0.48\\
O & 2006-Mar-02/12:00  &0.93$\times$0.30; -3.4$^{\circ}$   & 20 & 0.46\\
P & 2006-Mar-17/11:02  &0.83$\times$0.26; -8.6$^{\circ}$   & 26 & 0.38\\
Q & 2006-Mar-30/10:05  &0.87$\times$0.28; -7.0$^{\circ}$   & 17 & 0.53\\
R & 2006-Apr-13/09:07  &0.78$\times$0.26; -9.2$^{\circ}$   & 30 & 0.53\\
\hline
\end{tabular}
\end{table*} 

\begin{figure}[!th]
   \centering
  \hspace{-0.7cm}\includegraphics[height=0.36\textwidth, trim=0 0 0 0, clip]{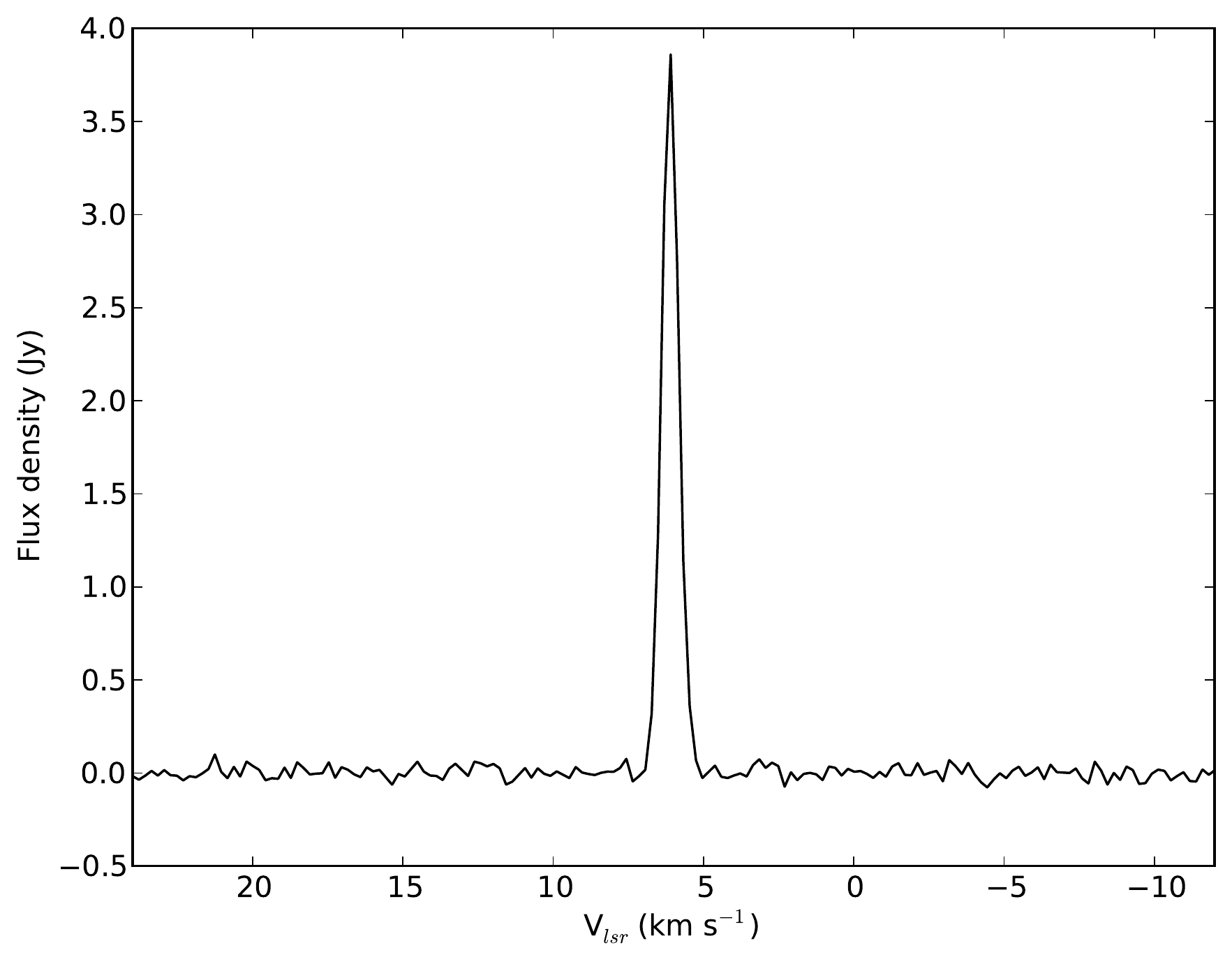} 
   \caption{ Observed spectrum of the water emission toward the brightest spot in epoch M (also called spot 2 in section 3.2).}
   \label{fig:spec}
\end{figure}

\begin{table*}[!ht]
\renewcommand{\arraystretch}{1.31}
\centering
\caption{Other water maser features detected toward I16293 with the VLBA.}
\label{tab:others}
\begin{tabular}{lcccccc}
&&&&\\ \hline
Epoch & Date of        &  channel & $\varv_{\rm LSR}$   & R.A. & Dec. &  Peak             \\ 
      & observation    &          & (km~s$^{-1}$)     &  ($^{\rm h}$ $^{\rm m}$ $^{\rm s}$)  & ($^\circ$ $'$ $''$)   &  (Jy~beam$^{-1}$) \\ \hline
F     & 05-Oct-13      & 268  & 2.8   & 16 32 22.864679(3) & -24 28 36.32705(5) & 0.31$\pm$0.03 \\
I     & 05-Nov-27      & 271  & 2.1   & 16 32 22.865514(1) & -24 28 36.33138(5) & 0.25$\pm$0.02 \\
J     & 05-Dec-08      & 271  & 2.1   & 16 32 22.865709(1) & -24 28 36.33135(2) & 1.05$\pm$0.03 \\
R     & 06-Apr-13      & 262  & 4.0   & 16 32 22.800719(1) & -24 28 36.57525(5) & 0.20$\pm$0.02 \\
R     & 06-Apr-13      & 272  & 1.9   & 16 32 22.882063(1) & -24 28 36.42483(4) & 0.27$\pm$0.01 \\
\hline
\end{tabular}
\end{table*}

The strong quasar 3C345 was observed at the beginning and at the end of each epoch and was used as a fringe finder. In total, three hours of observation were spent on I16293 during each observation\footnote{The source YLW16A was also observed in each epoch, but these data are not discussed in the present work.}, which was correlated at position $\alpha_{\rm J2000} = 16^{\rm h}32^{\rm m}22\rlap{.}^{\rm s}889$ and $\delta_{\rm J2000} = -24^{\circ}28'36\rlap{.}''25$. 

Data calibration and imaging were performed using the Astronomical Image  Processing System \citep[AIPS;][]{Greisen_2003}. First, we removed the delays introduced by the ionospheric content; then we applied corrections to the Earth Orientation Parameters used by the correlator, and corrections for voltage offsets in the samplers. Instrumental single-band delays were then determined and removed using fringes detected on 3C345.  Amplitude calibration was done using the provided gain curves and system temperatures to derive the System Equivalent Flux Density (SEFD) of each antenna. At this stage, the bandpass shape calibration, which was obtained from the scans on 3C345, was applied to the data.   In order to set the velocity information for I16293, we used the radio velocity definition and adopted a line rest frequency of 22.23508~GHz. Then, we  corrected the data  for the Doppler  shift due to the rotation of the antennas with the Earth around the Sun during observations.  Finally, global fringe fitting was run on J1625-2527 in order to find residual phase rates and delays; the solutions from this final step were then applied to the maser line data. Due to poor weather  conditions or technical faults at some antennas of the array, the  first three epochs presented very poor data quality and were discarded from the analysis. 

\begin{figure*}[!ht]
   \centering
  \includegraphics[height=0.99\textwidth,trim= 5 20 0 15, clip]{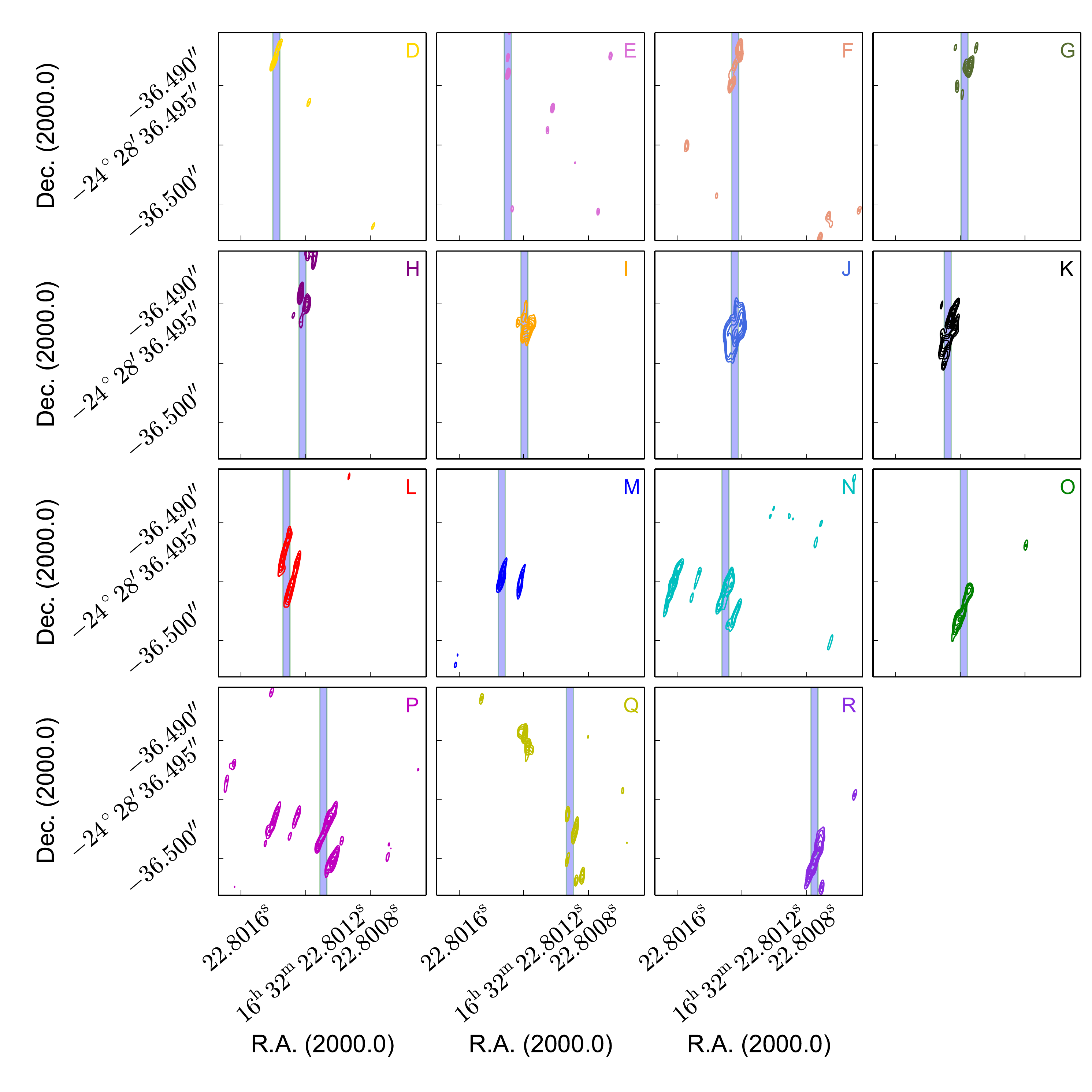}
   \caption{Water masers detected in our VLBA images above $10\sigma$ at each epoch.  The images show the emission from the channel at $\varv_{\rm LSR}=6.1$~km~s$^{-1}$. Contour levels are 0.3, 0.4, 0.5, 0.65, 0.80, and 0.95 times the peak level on the images, which are given in Table~\ref{table1}. The blue strips indicate the R.A. position with error bars of the shock front (see Section 3.2.)  }
   \label{fig:full}
\end{figure*}

We found strong maser emission at an LSR velocity, $\varv_{\rm LSR}$, of $\approx 6.1$~km~s~$^{-1}$ (channel 252; see Figure \ref{fig:spec} for one example) in all the 15 remaining epochs.  Images were produced  for this velocity channel (c.f.\ Sec. \ref{sec:prop}) using a pixel size of 50~$\mu$as  and a weighting scheme intermediate between natural and uniform (ROBUST = 0 in AIPS).
The emission from this channel was found to be offset from the phase center by $\sim1\rlap{.}''2$. Therefore the phase center was shifted (using the task UVFIX in AIPS) to the position of the emission and then used to produce our final images, shown in Figure~\ref{fig:full}.  The resulting {\it r.m.s.} noises in these images are given in Table \ref{table1}.


The final images were exported to FITS format for further analysis using the CASA software package. This software provides more flexibility for fitting multiple Gaussians in regions containing a large number of pixels (task IMFIT in CASA versus task JMFIT in AIPS). The Gaussian fittings were used to determine the position and flux of the water masers spots. We compared the fitted positions from both softwares and found that both agree within 1$\sigma$. JMFIT and IMFIT provide an estimate of the position errors based on the expected theoretical astrometric precision of an interferometer (Condon 1997). 

The  barycentric coordinates of the Earth, as well as the Julian date of each observation, which are necessary for the astrometric fits, were calculated using the NOVAS routines distributed by the US Naval Observatory.

\section{Results}

\subsection{Structure and properties of the emission}\label{sec:prop}
Maser emission is detected at radial velocities $\varv_{\rm LSR}  \sim 1.9$, 2.1, 2.8, 4.0, and 6.1~km~s$^{-1}$.  These velocities are very similar to the systemic LSR velocity of I16293 A, for which \cite{jorgensen2011} give +3.2 km~s$^{-1}$. Only the emission at 6.1~km~s$^{-1}$ is detected in all of the 15 epochs, while emission at the other velocities is not seen in more than two epochs (See Table \ref{tab:others}). For this reason, we  focus our analysis on the persistent emission at 6.1~km~s$^{-1}$.  We note that \cite{imai2007} also detected emission in their VERA observations from 2005 and 2006, at a velocity of $\varv_{\rm LSR}~=~6.0$~km~s$^{-1}$ with S$_\nu\sim 2$ Jy, which may correspond to the emission detected here.

For the astrometric analysis we only consider the spots detected in the maps for this velocity channel (channel 252). 
In general, several spots are detected at each epoch. Some spots show an elongated structure, just as was previously reported by \cite{imai2007}. As noticed by these authors,  the brightness distribution of the emission is spatially resolved and variable. The  brightest spot at one epoch is not necessarily the brightest one in the other epochs and many of them were detected in one epoch only.  For all these reasons, choosing the correct positions  for the astrometry implies a careful analysis of the images.

\subsection{Selection of positions for astrometry}\label{sec:selection}

\begin{figure*}[!th]
   \centering
   \begin{tabular}{cc}
  \includegraphics[height=0.4\textwidth,trim= 0 0 0 0, clip]{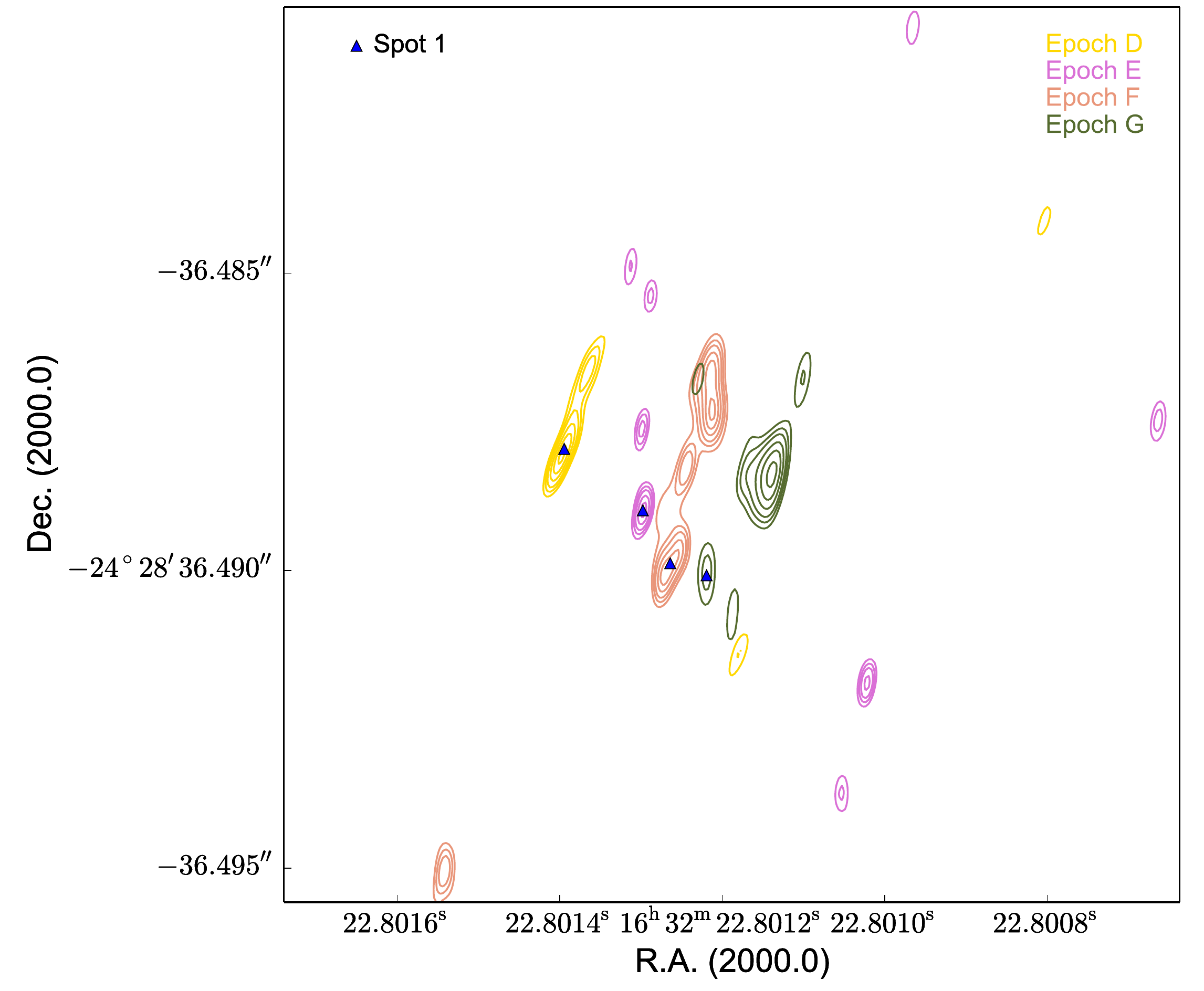}  & \includegraphics[height=0.4\textwidth,trim= 0 0 0 0, clip]{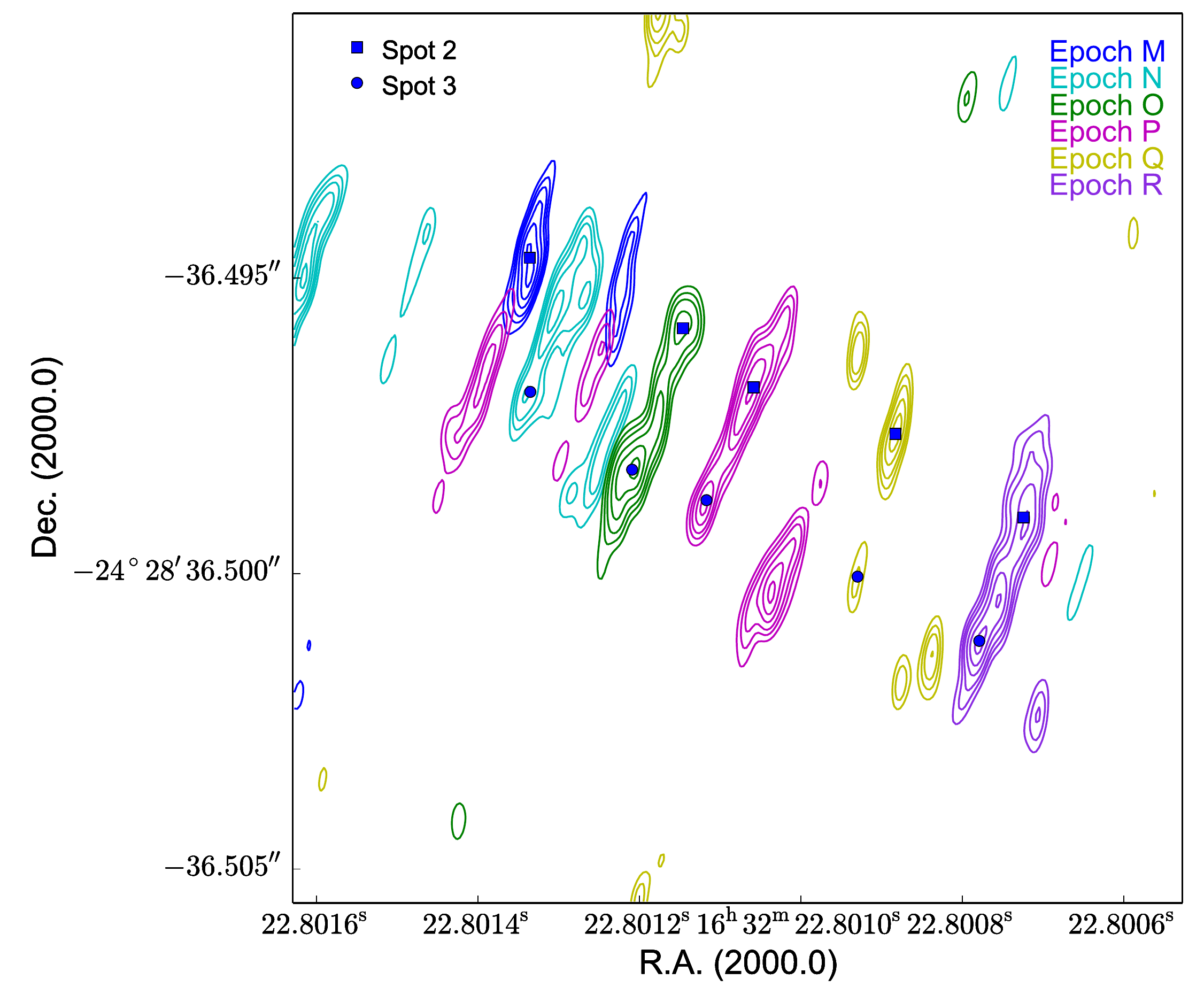}
  \\
  \end{tabular}
   \caption{Superposition of the detected spots. For clarity and to guide the eye, the left panel shows spots detected in the first four epochs (epoch D to G; c.f.\ Table~\ref{table1}). The right panel shows spots detected from epochs M to  R. The contour levels are 0.3, 0.4, 0.5, 0.65,0.80, and 0.95 times the peak flux of each image. Triangle symbols indicate the emission peaks of spot 1. The squares and circles correspond to spots 2, and 3, respectively. }
   \label{fig:sspots}
\end{figure*}

In order to perform accurate and reliable astrometry, the position of the same object  needs to be measured at several consecutive epochs.
For compact objects, such as stellar non-thermal continuum sources, there is little or even null confusion of the source position from epoch to epoch: it is the position of the emitting star. However, the case of water masers in nearby star forming regions is much more complex. 
Due to their variability and motions, it is not easy to  associate different spots even between contiguous epochs, especially when the emission is heavily resolved. 
The water maser shocks are located at the base of YSOs outflows \citep{torrelles2003,sanna2012}. Their motion on the plane of the sky are the  result from the superposition of a proper motion plus a reflex motion introduced by the parallax. Especially important is their movement in the Right Ascension (R.A.) direction, since their position in this direction is better determined and allow better determinations of the trigonometric parallax \citep{reid2009}.

To select the spots we made the following assumptions: (i) all the spots lie at the same distance, (ii) they have similar 
proper motions, and (iii) their radial velocity is constant at the different epochs. According to these assumptions, 
we expect that the position angle and projected angular distance between two different spots do not significantly change 
in a time interval of 15 days. With this in mind, we can identify individual spots between consecutive epochs as shown in 
Figure~\ref{fig:sspots}. In the left panel, we superpose the maser emission detected in epochs D-G, which were observed 
from September to November,  2005. We can clearly identify a spot that is detected in these four epochs and that shows a 
coherent motion toward the south-west of the map. We will refer to this spot as ``spot 1''. In the right panel of 
Figure~\ref{fig:sspots}, we now show the detected emission in epochs M-R, which were taken from January to April, 2006. 
In these epochs, the structure of the emission is much more complex. We can still distinguish two spots with coherent
motions also toward the south-west. All of these are detected in epochs O, P, Q and R. These two spots show a similar
position angle and angular separation (of about $2.5$~mas) in the four epochs. We call them ``spot 2'' and ``spot 3'' 
(see Figure \ref{fig:sspots}). Spot 2 is also detected in epoch M, but not in epoch N, and, conversely, spot 3 is  detected 
in epoch N but not in epoch M.   We discarded epochs H to L because the maser emission shows multiple blended peaks, which cannot be distinguished unambiguously.  The positions of spots 1, 2, and 3 are listed in Table~\ref{tab:pos} and they will be used in Section 3.3 to estimate the parallax of I16293.

\begin{table}[!htbp]
\renewcommand{\arraystretch}{1.31}
\small
  \begin{center}
  \caption{Detected epochs and positions of the different maser spots used for the astrometry.}
    \begin{tabular}{ccccc}\hline\hline
JD & $\alpha$  ($^{\rm s}$)             & $\sigma_{\alpha}$  & $\delta$  $('')$     & $\sigma_\delta$  \\
   &$16^{\rm h}32^{{\rm m}}$		& $\times 10^{-7}$ ($^{\rm s}$)       & $-24^{\circ}28'$   &   $\times 10^{-5}$ $('')$     \\
\hline
   \multicolumn{5}{l}{(Spot 1)}\\
\hline
2453626.46& 22.8013945 &8 &36.48795& 3 \\
2453639.43& 22.8012977 &4 &36.48899& 2 \\
2453657.36& 22.8012641 &9 &36.48988& 3 \\
2453676.33& 22.8012193 &5 &36.49008& 2 \\
\hline
   \multicolumn{5}{l}{(Spot 2)}\\
\hline
2453754.12 & 22.8013360&  6&  36.49466&  2\\
2453797.00 & 22.8011462&  3&  36.49585&  2\\
2453811.96 & 22.8010585&  7&  36.49685&  3\\
2453824.92 & 22.8008828&  3&  36.49764&  2\\
2453838.88 & 22.8007241&  6&  36.49905&  2\\
\hline
   \multicolumn{5}{l}{(Spot 3)}\\
\hline
2453771.07&  22.8013353&  9&  36.49693&  3\\
2453797.00&  22.8012091&  6&  36.49824&  3\\
2453811.96&  22.8011170&  6&  36.49876&  3\\
2453824.92&  22.8009297&  4&  36.50005&  2\\
2453838.88&  22.8007789&  6&  36.50114&  2\\
    
\hline
    \label{tab:pos}
    \end{tabular}
  \end{center}
\end{table}


 It is clear from this analysis that the spots are very variable and that their associated emission lasts for only a few months. A second approach for the determination of the trigonometric parallax is to track the motion of the shock front, which is traced by the collection of several spots at each epoch, and then fit the positions of this shock front in the R.A. direction. We adopted the mean position of the spots associated with it, which were identified by their proximity in the maps, as the position of the shock front at each epoch. These positions are listed in Table~\ref{tab:posFS} and shown as blue strips in Figure~\ref{fig:full}. In the epochs where we see a lot of structure in the emission (for example in epochs N and P) we only consider the spots at the expected positions according to the advancing motion of the shock front. 

Having identified three main spots that persisted in at least four epochs {and the mean positions of the shock front}, we now proceed with the astrometric fits. For this purpose, we apply the single value decomposition (SVD) fitting  scheme described by \citet[][]{loinard2007}.

\begin{table}[h!]
\centering
\small
\renewcommand{\arraystretch}{1.31}
\caption{Observed epochs and R.A. positions of the water maser shock front. }
\begin{tabular}{cccc}
&&&\\ \hline
    &Julian     & R.A. (s)                 & $\sigma_{\rm R.A.}$  \\ 
Epoch&Date       & 16$^{\rm h}$32$^{\rm m}$ &$\times 10^{-6}$ ($^{\rm s}$)  \\
\hline
D&2453626.46 & 22.801381  & 2 \\
E&2453639.43 & 22.801298  & 1 \\
F&2453657.36 & 22.801242  & 8 \\
G&2453676.33 & 22.801173  & 6  \\
H&2453687.30 & 22.801209  & 4  \\
I&2453702.26 & 22.801206  & 4 \\
J&2453713.23 & 22.801244  & 4 \\
K&2453727.19 & 22.801277  & 5 \\
L&2453742.15 & 22.801318  & 5 \\
M&2453754.12 & 22.801336  & 4 \\
N&2453771.07 & 22.801303  & 10 \\
O&2453797.00 & 22.801178  & 2 \\
P&2453811.96 & 22.801090  & 6 \\
Q&2453824.92 & 22.800915  & 3 \\
R&2453838.88 & 22.800752  & 8 \\
\hline
\label{tab:posFS}
\end{tabular}
\end{table}

\begin{table*}[!h]
\small
\renewcommand{\arraystretch}{1.31}
  \begin{center}
  \caption{Parameters obtained from the astrometric fits. The systematic errors that were added quadratically 
     are also presented.  }
    \begin{tabular}{cccccccccc|cc}\hline\hline
Fit & Spot  & $\pi\pm\sigma_{\pi}$ & $d$ &$\mu_{\alpha}\cos{\delta}\pm\sigma_{\mu_{\alpha}\cos{\delta}}$ &
$\mu_{\delta}\pm\sigma_{\mu_{\delta}}$&\multicolumn{2}{c}{ Offsets}& \multicolumn{2}{c|}{Post-fit rms}&\multicolumn{2}{c}{Systematic Errors}\\
    &  &           (mas)      &  (pc)   & (mas yr$^{-1}$)  & (mas yr$^{-1}$)&\multicolumn{2}{c}{ (mas)}&\multicolumn{2}{c|}{(mas)}   &\multicolumn{2}{c}{(mas)}\\
    &        &         &           &          &            &$\alpha$   &   $\delta$ &  $\alpha$   &   $\delta$       & $\alpha$     &   $\delta$     \\
\hline
\multirow{3}{*}{Individual}& 1    &  $6.9\pm4.7$  &145$^{+314}_{-59}$&$-44\pm19$&$-13\pm6$&...&...&0.17&0.32&0.33&0.63\\
&2    &  $7.1\pm1.3$  &$141^{+30}_{-21}$ &$-39\pm2$&$-16\pm5$ &...&...&0.21&0.54&0.33&0.84\\
&3    &  $8.3\pm2.1$  &$120^{+40}_{-24}$ &$-37\pm3$&$-21\pm4$ &...&...&0.24&0.34&0.36&0.52\\ \hline
\multirow{3}{*}{Simultaneous}&1   &  \multirow{3}{*}{$7.4\pm1.1$}  & \multirow{3}{*}{$135^{+23}_{-16}$} & $-47\pm5$&$-13\pm6$&...&...&0.17 & 0.32 & 0.33 & 0.62 \\
&2   &   &    & $-40\pm2$&$-16\pm5$& ...&...& 0.22 & 0.54 & 0.33 & 0.84 \\
&3   &   &    &$-38\pm3$&$-21\pm4$& ...&...&  0.25 & 0.33 &  0.39 & 0.52 \\\hline
\multirow{2}{*}{With offset}&1   &  \multirow{2}{*}{$7.1\pm1.3$}  & \multirow{2}{*}{$141^{+30}_{-21}$} & $-42\pm3$&$-13\pm8$&$0.0\pm0.0$&$0.0\pm0.0$&0.17 & 0.32 & 0.36 & 0.80 \\
&2   &   &    & $-36\pm2$&$-16\pm5$&$-0.4\pm1.0$&$1.4\pm2.3$& 0.22 & 0.54 & 0.36 & 0.80 \\ \hline
{Spots 1 \& 2 }& 1+2  &  7.1$\pm$0.5  &141$^{+10}_{-9}$  &$-39\pm2$&$-13\pm1$ &...&...&0.29&0.48&0.36&0.58\\
\hline
 Shock front &   --     & $6.6\pm0.6$ & $152^{+16}_{-12}$ & $-32\pm3$ &...& ...&...&0.14 & ... & 0.64 & ...\\ \hline
\hline
    \label{tab:ffit}
    \end{tabular}
  \end{center}
\end{table*}

\begin{figure*}[!t]
   \centering
   \begin{tabular}{cc}
  \hspace{-0.7cm}\includegraphics[height=0.36\textwidth, clip]{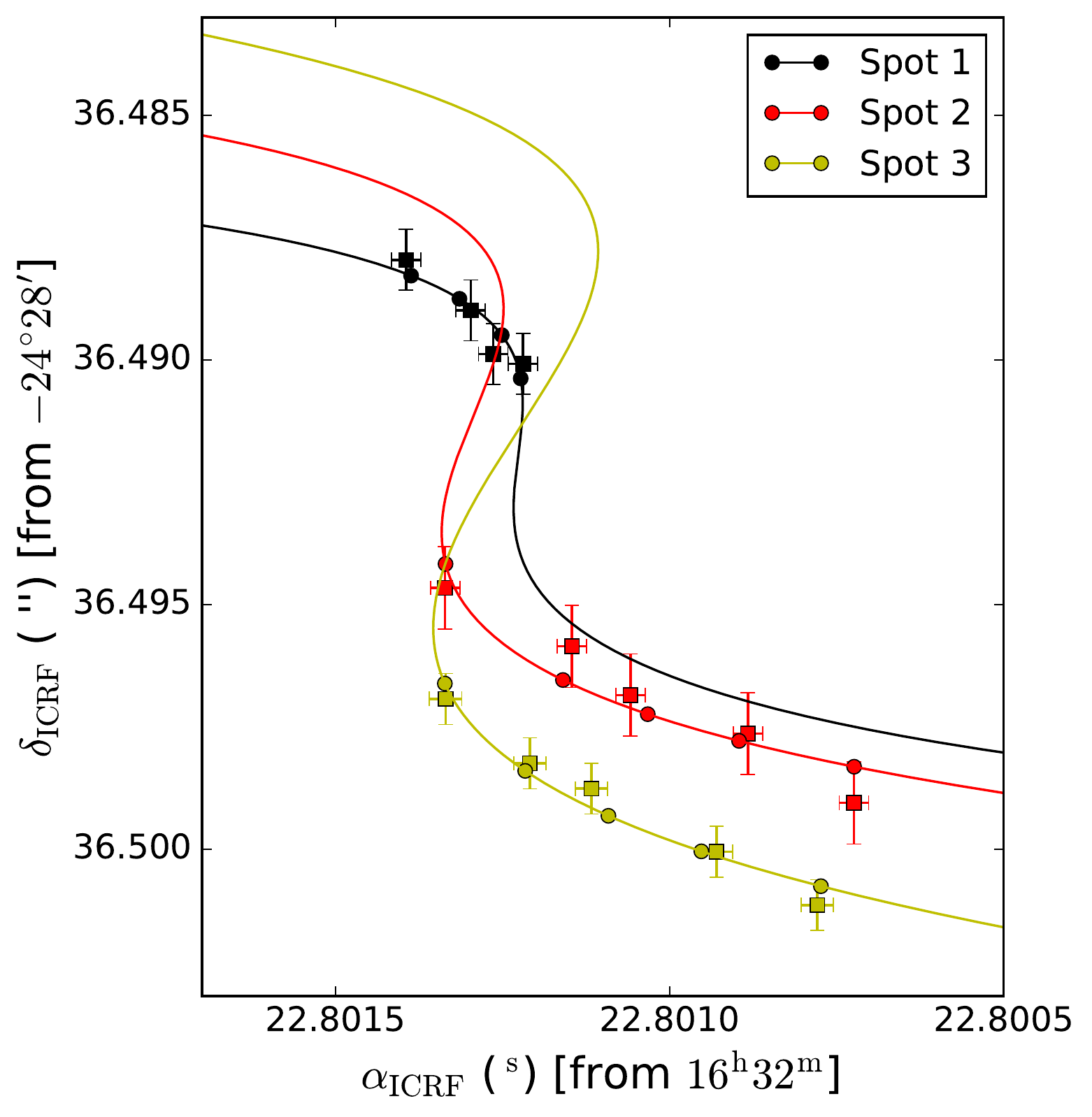}  & 
  \hspace{-0.4cm}\includegraphics[height=0.36\textwidth, clip]{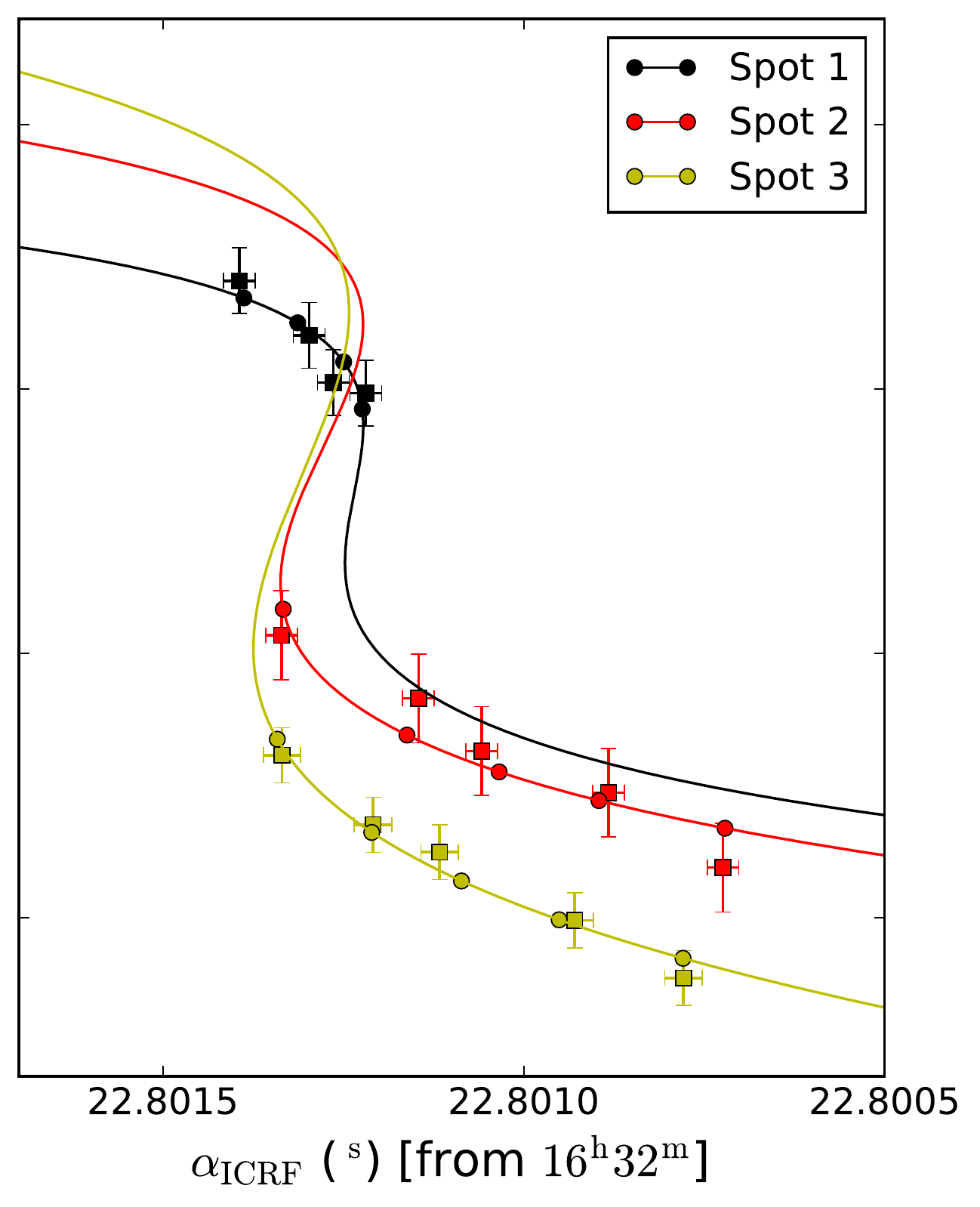}\\
  \hspace{-0.7cm}\includegraphics[height=0.36\textwidth, clip]{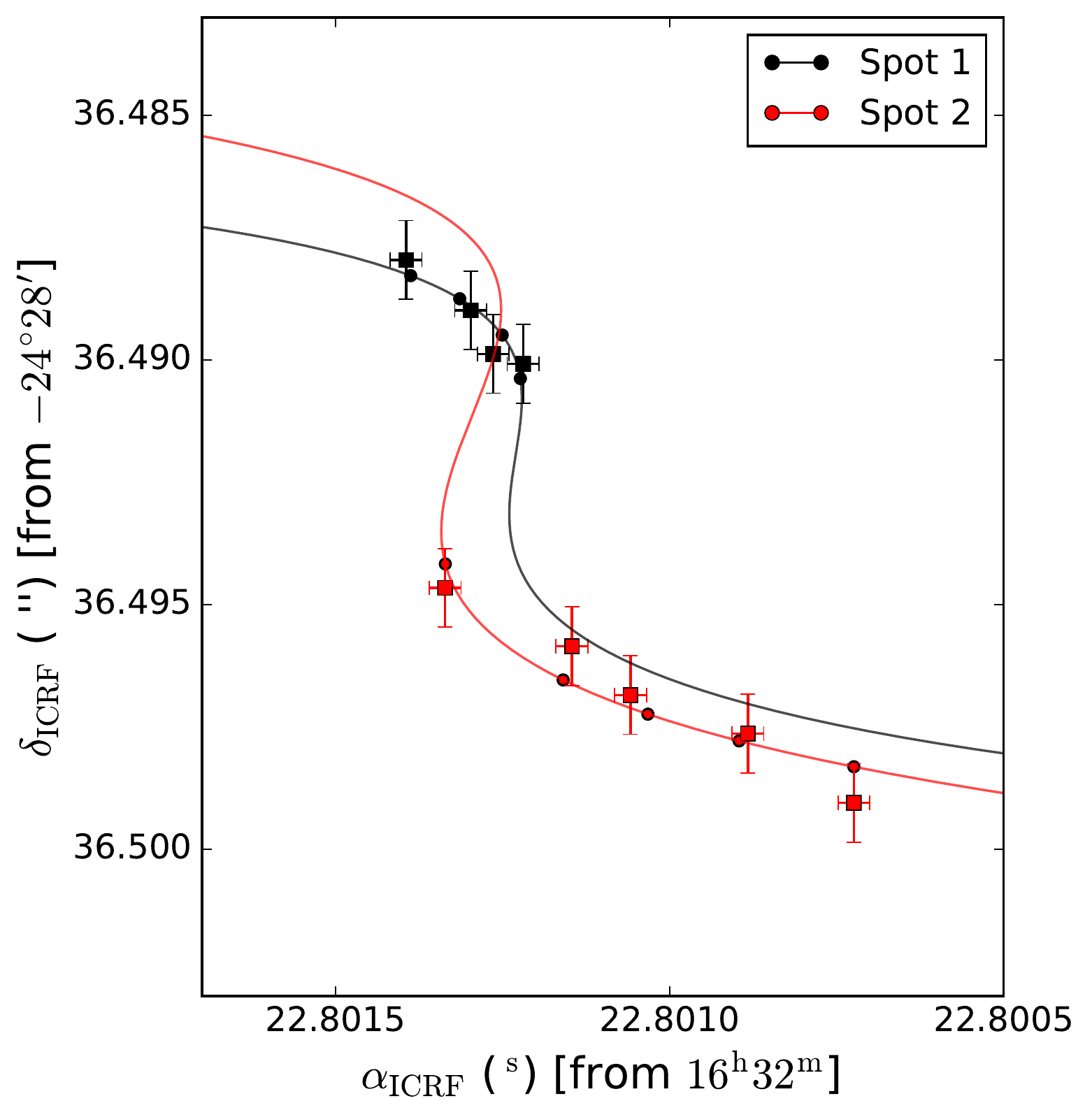}  & 
  \hspace{-0.4cm}\includegraphics[height=0.36\textwidth, clip]{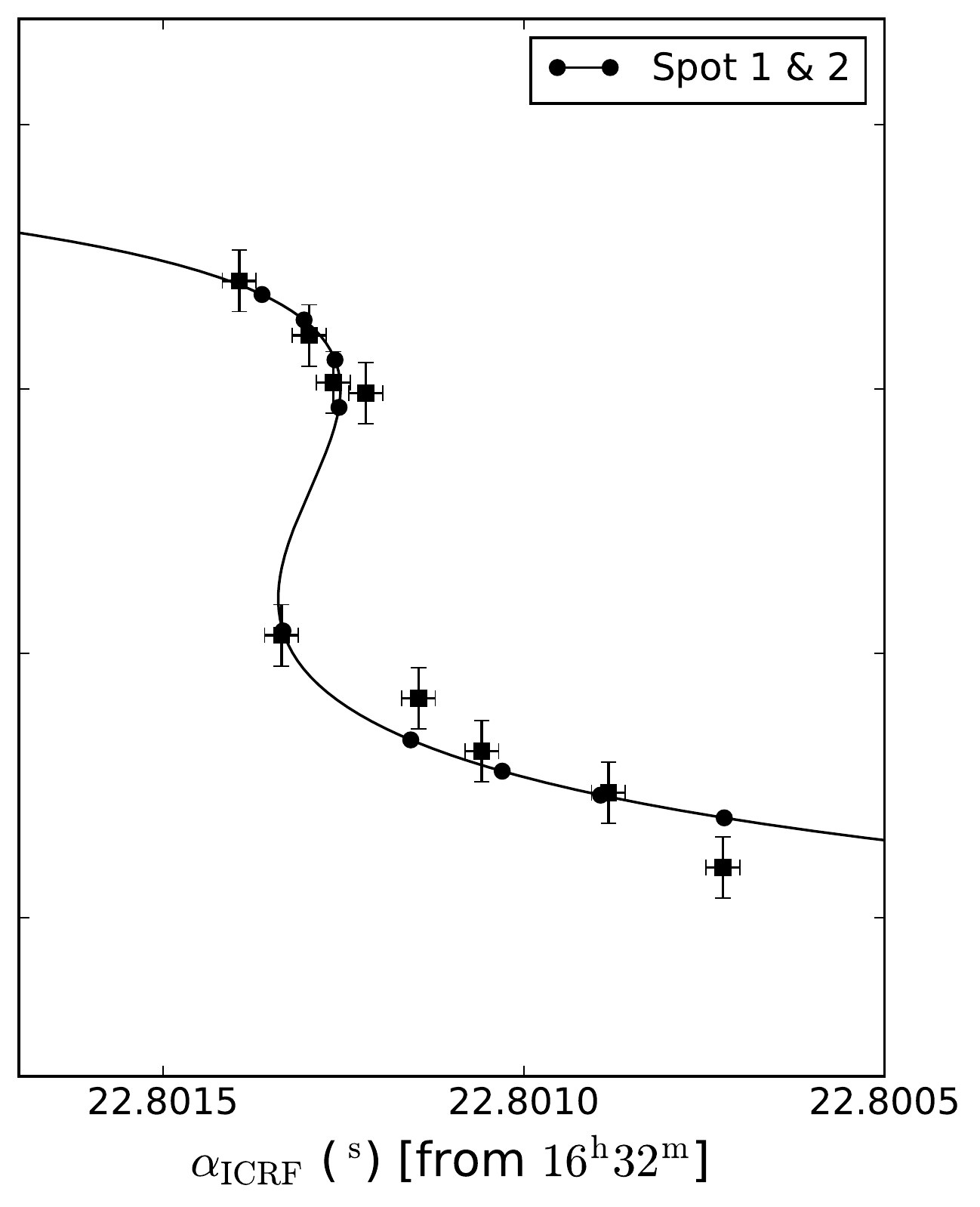}
  \end{tabular}
   \caption{Resulting fits to the water maser spots detected toward I16293, see 
   Table \ref{tab:ffit}. The black, red, and yellow lines show the best fits for 
   spot 1, 2 and 3, respectively. The solid  squares with the error bars indicate  
   measured positions and their corresponding uncertainties. The solid circles mark 
   the expected positions from the fits at the observing epochs. 
   {\it Top left}: Individual fits to each of the three maser spots. 
   {\it Top right}: Simultaneous fit  assuming that the parallax is the same for 
   the three spots. {
   {\it Bottom left}: Simultaneous fit to positions of Spot 1 and 2, by allowing them to have a positional offset and different proper motions.   
   {\it Bottom right}: Simultaneous fit to spots 1 and 2, assuming that they belong to the same cloudlet.}
   }
   \label{fig:fits}
\end{figure*}

\subsection{Astrometry}

In radio astrometry, the source positions are commonly registered at multiple epochs spanning long periods, usually one year 
duration in order to trace the full parallax sinusoid. The observations presented here cover only a time period of seven months.
However, the first and last data were taken close to the time when the maximum parallax angle is reached, i.e., in September 
and March.  Radio astrometry has been performed with similar time spans for other star forming regions with good determination 
of parallaxes \citep[e.g.,][]{hirota2008,kim2008}. As described in Section \ref{sec:selection}, only three spots are detected in four or five epochs each (see Table~\ref{tab:pos}), with time spans of only 2 to 3 months. The shock front, on the other hand, is tracked over the full seven months. We can compare the astrometry derived from the individual spots to that derived from the positions of the shock front. 

We first fit separately each of the three maser spots (Figure~\ref{fig:fits}, first panel). The resulting values of the  trigonometric parallax and distance from these fits are given  in Table~\ref{tab:ffit}. Since the results from the three fits are consistent with each other within 1$\sigma$, we simultaneously fit the three maser spots by requiring a single parallax, and allowing the proper motions to vary between the different spots. The parallax obtained in this case is $7.4\pm1.1$ mas, corresponding to a distance of $135^{+23}_{-16}$ pc. The values of the proper motions do not significantly change when compared to 
those derived from the individual fits.

We note that, if we extrapolate the positions of spot 1 to the epochs when spot 2 is detected (by using the best models from the previous two fits), then their corresponding positions are consistent within the errors. 
This suggests that these two spots are related and may indeed correspond to the same gas condensation 
\cite[or cloudlet;][]{sanna2017}. To test this hypothesis, we perform a simultaneous fit to both 
spots and consider that there is a positional offset between them.  The resulting astrometric parameters
from this fit are also given in Table~\ref{tab:ffit} and shown in Figure~\ref{fig:fits} (Bottom-left). The best-fit parallax yields a distance of 141$^{+30}_{-21}$ pc. It is important to mention that the resulting positional offset is relative small and the proper motions of the spots are consistent within errors (see Table 5).
Thus, an additional fit was done by assuming that spots 1 and 2 trace the motion of the same cloudlet.
 From this fit, we found that the parallax uncertainty is at least a factor of two lower than the obtained in the previous fits. However,  given the strong variability of the maser emission, the spots could be bright at different times, and thus we consider the fit with non-zero positional offset to be more robust.

\begin{figure}[!t]
   \centering
  \hspace{-0.7cm}\includegraphics[height=0.42\textwidth, trim=60 20 0 0, clip]{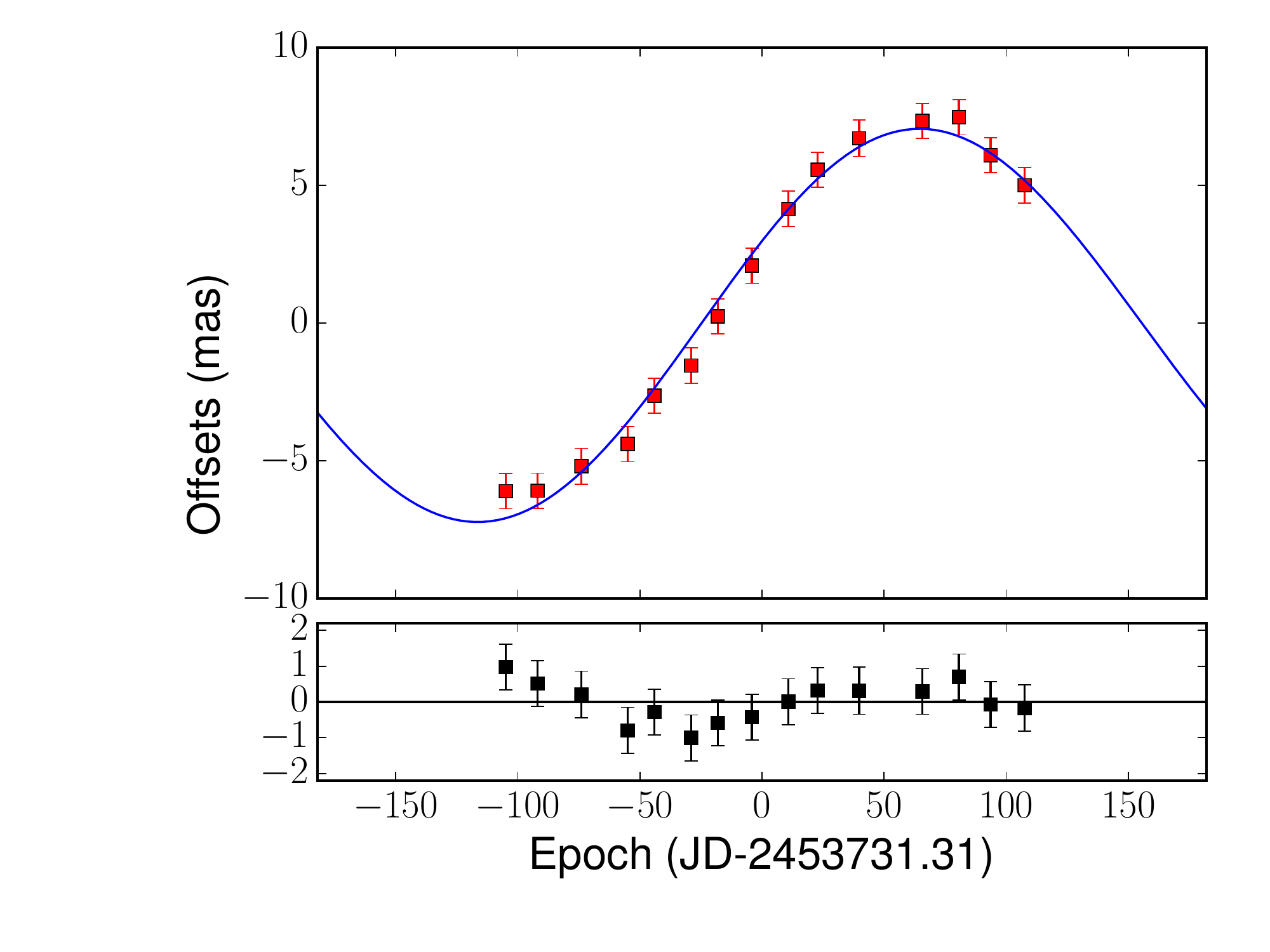} 
   \caption{ Parallax fit to the mean positions of the shock front. {\it Top:} The blue line is the parallax sinusoid from the best fit to these data. The red squares are the mean positions of the shock front. In both cases, we have subtracted the proper motion and the R.A.\ position at JD 2453731.31 (the mean epoch from our observations) from the best fit. {\it Bottom}: Residuals in right ascension. 
}
   \label{fig:fitRA}
\end{figure}

Second, we performed an astrometric fit to the mean position of the shock front and derive a trigonometric parallax of  $\pi=6.6\pm0.6$~mas, corresponding to a distance of d\,$=152^{+16}_{-12}$~pc. We show the parallax sinusoid from this fit in Figure~\ref{fig:fitRA}, and give the resulting parameters in Table~\ref{tab:ffit}. The values of the parallax derived from the  fits to the three main spots are in good agreement with that obtained from the fit to the R.A.\ motion of the shock front. 

 The correlation coefficient matrices, for all fits, are shown in Appendix~\ref{sec:CM}. From these, we noticed a high (anti-)correlation between the parallax and the other fitted parameters in all the fits, as is expected given the short timescale covered by the observations. To check the reliability of our results, in view of  these high correlations, we performed a series of Monte Carlo simulations which are presented in Appendix~\ref{sec:MC}. 
The result from these simulations is that, given the statistical and systematic errors in our measurements,  and even when the observations cover a period of only 3--7 months, the true astrometric parameters can be recovered from our fits. The worst case is the individual fit to spot 1, because this spot was only detected in 4 epochs and, consequently, the uncertainties on the fitted parameters are significantly larger.

\section{Discussion}

\subsection{The distance to I16293}

As we mentioned in the introduction, \cite{imai2007} obtained a distance of 
178$^{+18}_{-37}$~pc to I16293 by measuring the trigonometric parallax of 
water masers associated  to this YSO. \cite{imai2007} mentioned 
that the most severe factor causing their modest astrometric accuracy was 
the temporal variation in the brightness structure of the maser spots they 
detected. If water masers have short  lifetimes of a few months, as is the 
case with water masers in I16293, it is possible to misidentify  maser spots
even from one epoch to the next. Taking into account this possibility, these 
authors performed several fits by using different combinations of observed 
positions. By excluding the oddest positions, they obtained an upper limit to
the trigonometric parallax of  7.1~mas (corresponding to a lower limit in the distance of $d=141$~pc). 
In the following, we discuss the differences between our analysis and that of \citet{imai2007}.

The first difference comes from their criterion for the selection of maser spots. 
They assumed that the brightest peak always traces the same 
spot, and used its measured positions for the astrometric fits. As we 
can see in Figure \ref{fig:sspots} this assumption is not valid, for instance from 
epoch M to epoch N. Additionally, the radial  velocities of their spots vary from 
one epoch to the other by $\sim1$~km~s$^{-1}$, suggesting that they are not tracing 
the same gas \cite[e.g.,][]{moscadelli2006}. 
Finally, the time intervals between their observed epochs range from 22 days up to 
three months, whereas our measurements are spaced, on average, by 15 days each one 
from the other. This higher cadence improves our ability to trace the same maser 
spots and, particularly, the shock front in time.

The estimated trigonometric parallaxes from all our fits agrees with the lower 
limit of 141~pc obtained by \cite{imai2007}. However, the fits that exhibit the smallest 
may be affected by the several assumptions we made regarding the associations between the 
detected spots (See sections 3.2 and 3.3). 
The most conservative result is thus the ’With offset’ fit (Table 5) since it allows the spots to have non-identical positions and also spans the largest period of observation (of seven months).  The resulting trigonometric parallax from 
this fit is $7.1\pm1.3$~mas, corresponding to a conservative value of $141^{+30}_{-21}$~pc for the distance.
This result also agrees within 1$\sigma$ with a recent value of  147.3$\pm$3.4~pc  measured 
to three other systems in Lynds 1689 \citep{OrtizLeon2017}.

\subsection{Relationship between the water maser emission and the molecular outflows in I16293}

\begin{figure*}[!t]
   \centering
   \begin{tabular}{cc}
  \hspace{-1cm}\includegraphics[height=0.38\textwidth,trim= 0 5 5 0, clip]{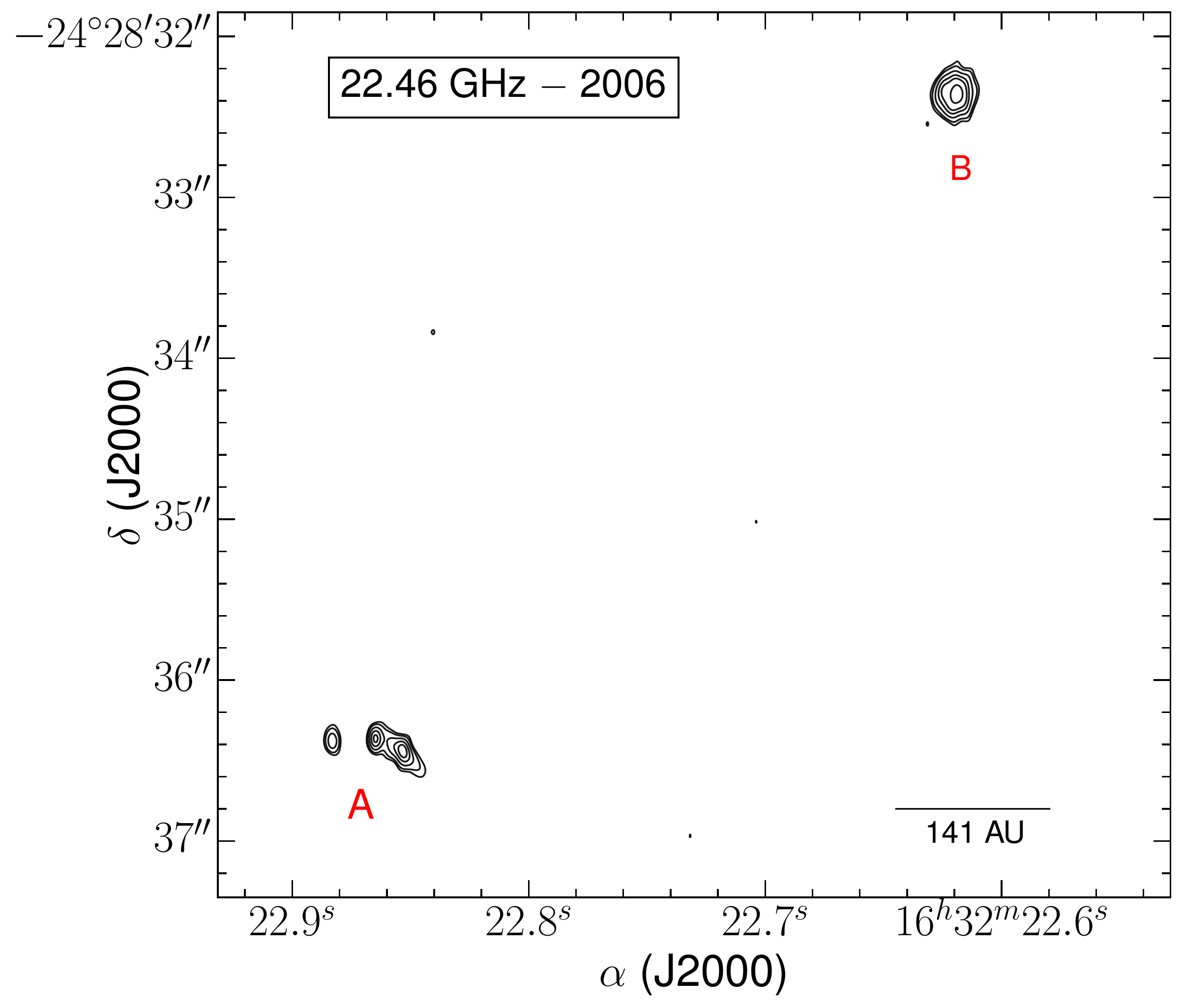}  & 
 \includegraphics[height=0.38\textwidth,trim= 0 5 5 0, clip]{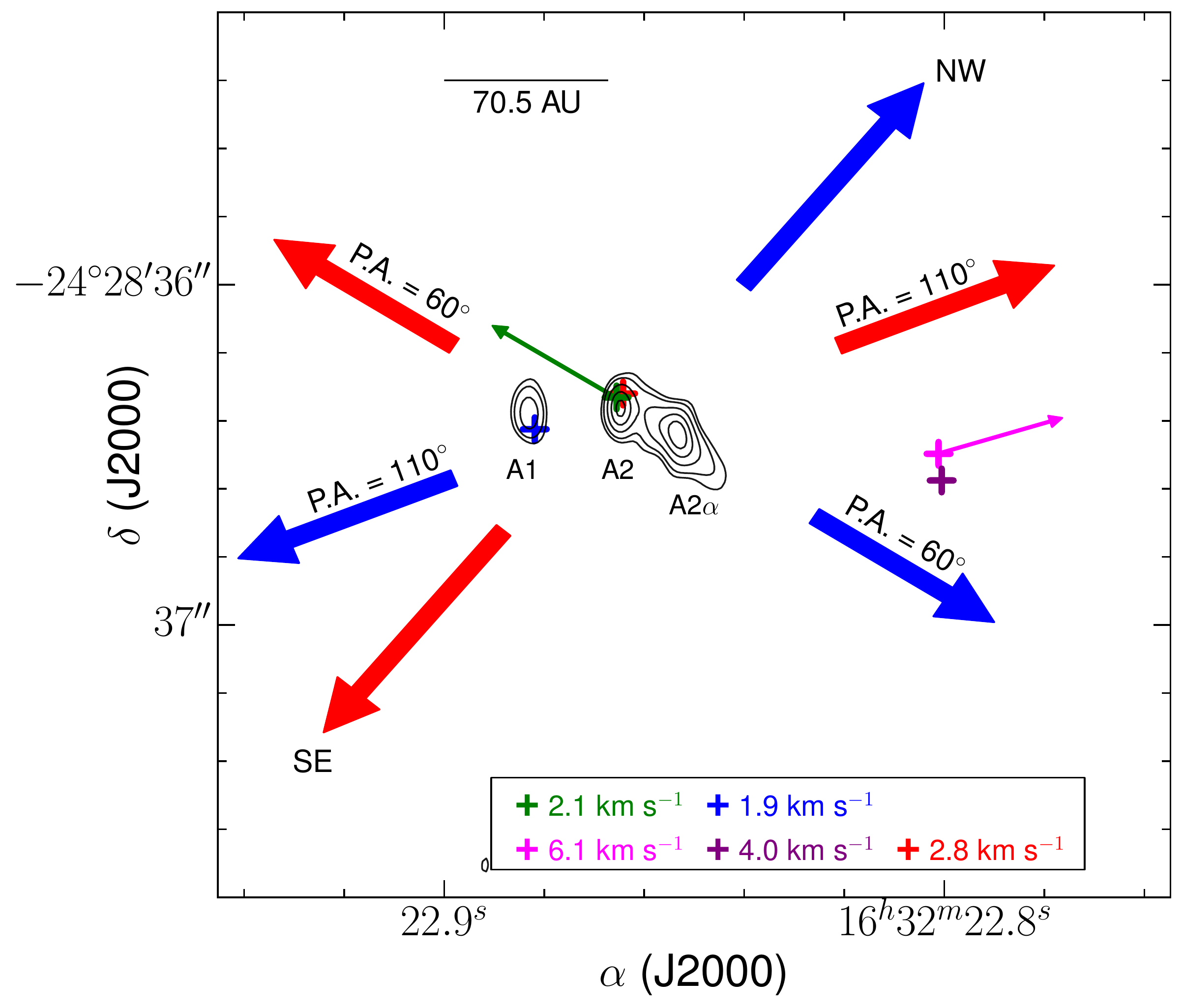} 
  \\
  \end{tabular}
   \caption{The I16293 system. {\it Left}: Continuum map at 22 GHz taken with the VLA on 2006 February 10. The noise level of this image is 47 $\mu$Jy beam$^{-1}$, and the contour levels are -4, 4, 6, 10, 15, 20, and 30 times this value. {\it Right}: Zoom on source A. We show the sources composing this system: A1, A2 and the ejecta A2$\alpha$. We also show the direction of the three outflows (blue and red arrows arising from this source). 
   The color crosses represent the position of the maser spots detected at different velocities (indicated with different colors; see Table \ref{tab:others}). The magenta cross indicates the position of the masers  at 6.1 km s$^{-1}$ 
and the magenta arrow arising from it shows the mean proper motion direction, while the green arrow indicates the proper motion direction of the maser spot at 2.1~km~s$^{-1}$ identified on epochs I and J. The blue cross indicates the position of the maser emission detected in epoch R at 1.9 km s$^{-1}$, very close to source A1, and the red cross marks the position of the maser spot at 2.8 km s$^{-1}$ close to source A2 and detected in epoch F. We also show the position of the maser spot detected in epoch R at 4.0 km s$^{-1}$ with a purple cross. None of the arrow lengths represent a velocity magnitude but only the direction of the outflows and the proper motions of the masers. We also include a reference of the physical scale on both images.}

   \label{fig:cont}
\end{figure*}

In Figure \ref{fig:cont} we show a radio continuum map of the I16293 system at  
22.46~GHz as observed with the VLA in 2006 February 10 (and previously reported 
by \citealt{pech2010}). This image shows the sources A and B that were introduced 
in section \ref{sec:intro}. On the right side of this figure we zoom in on 
source~A and show its two components A1 and A2, and the A2$\alpha$ ejecta as well. 
In Figure~\ref{fig:cont} the magenta cross shows the position of the persistent 
water maser peak observed at $\varv_{\rm LSR}=6.1$~km~s$^{-1}$. 

The proper motions of the continuum sources A1 and A2 are 
($-3.0\pm0.4$, $-27.9\pm0.8$)~mas~yr$^{-1}$ and 
($-7.8\pm0.6$, $-21.8\pm0.5$)~mas~yr$^{-1}$, respectively 
(Hern\'andez-G\'omez et al. in prep.). To convert proper 
motions into transverse velocities we use the relation 1~mas~$\equiv$~0.67~km~s$^{-1}$,
which is appropriate for a distance of 141~pc.
Here, we have measured a proper motion of ($-39\pm2$,~$-13\pm1$)~mas~yr$^{-1}$ 
for the persistent maser cloudlet (Spot 1 \& 2). The proper motion in declination of 
this cloudlet is similar in magnitude and direction to the one of source A2,
but significantly larger in the right ascension direction. After 
subtracting the velocity vector of source A2, we found that the intrinsic 
proper motion of the spot  is
($-31.2\pm2.1$,~$8.8\pm1.1$)~mas~yr$^{-1}$~=~($-20.9\pm1.4$,~$5.9\pm0.7$)~km~s$^{-1}$, 
i.e. toward the northwest of the continuum source. It is indicated with 
the magenta arrow in Figure \ref{fig:cont}. 

Another maser spot at a radial velocity of 2.1~km~s$^{-1}$ was detected in the consecutive epochs I and J and is indicated in Figure~\ref{fig:cont} with a green cross. We measured its positions and corrected for the trigonometric parallax in order to arrive at a proper motion of  ($43.6\pm0.9$, $7.5\pm1.9$)~mas~yr$^{-1}$. Thus, its relative proper motion with respect to source A2 is ($51.4\pm0.9$,~$29.3\pm2.0$)~mas~yr$^{-1}$~=~($34.5\pm0.7$,~$19.7\pm1.3$)~km~s$^{-1}$, i.e. toward the northeast of the continuum source, as it is indicated by the green arrow in Figure \ref{fig:cont}. The maser spots at other radial velocities were only detected once, and their positions are shown in Figure \ref{fig:cont} as crosses of different colors. We note that two of them are close to source A2 and the other is close to source~A1.

As mentioned before, several molecular outflows have been observed towards I16293 \cite[see e.g.,][]{mizuno1990,loinard2013,girart2014}. There are two bipolar outflows arising from source A with position angles (P.A.) of 60$^{\circ}$ and 110$^{\circ}$ approximately. There is also a third outflow in the SE-NW direction extending between sources A and B. The red- and blue-shifted lobes have radial velocities between  6 and 20~km~s$^{-1}$, and $-10$ and 2~km~s$^{-1}$, respectively \citep{mizuno1990}.
The direction of the lobes of these three outflows are also indicated in 
Figure \ref{fig:cont}. The water masers in I16293 are at the starting point 
of these much larger scale outflows.
Given that the water maser shocks are usually found at the base of YSOs 
outflows \cite[][]{torrelles2003,sanna2012}, we suggest that the maser emission at 2.1~km~s$^{-1}$, and at 6.1~km~s$^{-1}$ is associated with these outflows. 
In particular, the radial velocity, position, and proper motions of the spot at 6.1~km~s$^{-1}$ suggest that this is directly related to the red-shifted lobe of the outflow with P.A.=110$^{\circ}$. The radial velocity of the maser spot at 2.1~km~s$^{-1}$  suggest that it may be associated with a blue shifted lobe, possibly with the outflow with P.A.=110$^\circ$.
However, the direction of its proper motions are similar to the red-shifted lobe  of the outflow with P.A.=60$^{\circ}$.  Thus, we cannot unambiguously associate the maser emission with a specific flow.

\section{Conclusions}

We have analysed 15 archival VLBA observations at 22.2~GHz to measure  
the trigonometric parallax of water masers associated with the YSO I16293, 
in the Ophiuchus molecular cloud, and determine its distance. In our analysis
the main source of uncertainty is the maser variability and the short time 
span covered by the observations.

After a careful analysis of the maps, we identified three maser spots, 
and fitted their displacement on the plane of the sky as due to a linear 
motion and the reflex motion of the trigonometric parallax.  Additionally, 
we have also tracked the position of the shock front and fitted its 
motion in the R.A.\ direction. {All our fits agree within the errors 
and suggest a distance around 140 pc. Our most conservative approach yields
a trigonometric parallax of $7.1\pm1.3$~mas, corresponding 
to a distance of $141^{+30}_{-21}$~pc.}

This distance estimate is smaller than that reported by \citet{imai2007} by 
nearly 40 pc, but the two results are consistent within 1$\sigma$.  Our value 
for I16293 distance is also  21 pc larger than the 120 pc obtained by Loinard et 
al.~(2008) based on two objects in L1688 (their results were later corrected by
\cite{OrtizLeon2017}, who derived larger distances for both objects). 
Furthermore, our measured distance is in good agreement with recent 
measurements  by \citet{OrtizLeon2017} of trigonometric
parallaxes of magnetically active YSOs in the Ophiuchus complex, 
and especially in the L1689 cloud at a  
distance of $147.3\pm3.4$ pc.  Thus our results support the widely held 
view that I16293 is a member of the L1689 cloud in Ophiuchus.

The radial velocities and direction of the relative proper motions of the
water masers (with respect to source A2), agrees with those of the outflow
emission previously reported in this system.  In the case of the water masers
with $\varv_{\rm LSR}=6.1$~km~s$^{-1}$ it was possible to associate them with 
the red-shifted lobe of the outflow with a P.A.~=~110$^\circ$.

\begin{acknowledgements}
We would like to acknowledge the referee Mark Reid for his comments and suggestions that
improve the manuscript.
G.-N.O.L acknowledges support from the Alexander von Humboldt 
Foundation in the form of a Humboldt Fellowship. A.H.-G. and 
L.L. acknowledges the financial 
support of DGAPA, UNAM (project IN112417), and CONACyT, M\'exico.
The Long Baseline Observatory is a facility of the National Science Foundation operated under cooperative agreement by Associated Universities, Inc.

\end{acknowledgements}


\bibliographystyle{aa}
\bibliography{references}

\onecolumn
\begin{appendix}

{
\section{Correlation matrices}\label{sec:CM}

The purpose of this appendix is to show the correlations
matrices for all the fits made to our data. In the following 
subsections we keep the name of the fits as they were 
introduced in Table~\ref{tab:ffit}.

\subsection{Individual}
Spot 1
\[
\bordermatrix{
&\alpha & \delta & \mu_\alpha\cos(\delta) &\mu_\delta & \pi\cr
\qquad\alpha& 1.00 & -0.21 & -0.55 &  0.16 &  0.68 \cr
\qquad\delta&-0.21 &  1.00 &  0.31 &  0.82 & -0.31 \cr
\mu_\alpha\cos(\delta)&-0.55 &  0.31 &  1.00 & -0.23 & -0.99 \cr
\qquad\mu_\delta& 0.16 &  0.82 & -0.23 &  1.00 &  0.24 \cr
\qquad\pi & 0.68 & -0.31 & -0.99 &  0.24 &  1.00  \cr
}
\] \\

\noindent Spot 2
\[
\bordermatrix{
&\alpha & \delta & \mu_\alpha\cos(\delta) &\mu_\delta & \pi\cr
\qquad\alpha          & 1.00 & -0.07 & -0.27 &  -0.07 & -0.93 \cr
\qquad\delta          &-0.07 &  1.00 & -0.01 &  -0.94 & 0.07 \cr
\mu_\alpha\cos(\delta)&-0.27 & -0.01 &  1.00 & -0.01 & -0.10 \cr
\qquad\mu_\delta      &-0.07 & -0.94 & -0.01 &  1.00 &  0.08 \cr
\qquad\pi             &-0.93 &  0.07 & -0.10 &  0.08 &  1.00  \cr
}
\] \\

\noindent Spot 3
\[
\bordermatrix{
&\alpha & \delta & \mu_\alpha\cos(\delta) &\mu_\delta & \pi\cr
\qquad\alpha          & 1.00 & -0.23 & -0.79 & -0.08 & -0.97 \cr
\qquad\delta          &-0.23 &  1.00 &  0.15 & -0.92 & 0.23 \cr
\mu_\alpha\cos(\delta)&-0.79 &  0.15 &  1.00 &  0.05 &  0.62 \cr
\qquad\mu_\delta      &-0.08 & -0.92 &  0.05 &  1.00 &  0.08 \cr
\qquad\pi             &-0.97 &  0.23 &  0.62 &  0.08 &  1.00  \cr
}
\] 

\subsection{Simultaneous}

\begin{small}

\[
\bordermatrix{
&\alpha[1] & \delta [1] & {\displaystyle \mu_\alpha\cos(\delta) [1]} &\mu_\delta [1] & \alpha [2] & \delta [2] & \mu_\alpha\cos(\delta) [2] &\mu_\delta [2] & \alpha [3] & \delta [3] & \mu_\alpha\cos(\delta) [3] &\mu_\delta [3]  &\pi\cr
\qquad\alpha [1]          & 1.00  & -0.11 & -0.74& -0.04& -0.18&  0.07&  0.70& -0.05&  0.79& -0.05&  0.07& -0.06& -0.88\cr
\qquad\delta[1]           &-0.11 & 1.00&  0.04& -0.95&  0.03& -0.01& -0.10&  0.01& -0.11& 0.01& -0.01&  0.01&  0.12\cr
\mu_\alpha\cos(\delta)[1] &-0.74&  0.04&  1.00&  0.01&  0.07& -0.03& -0.28&  0.02& -0.32& 0.02& -0.03&  0.02&  0.35\cr
\qquad\mu_\delta [1]   &-0.04& -0.95&  0.01&  1.00&  0.01& -0.00& -0.03&  0.00& -0.04&  0.00&  0.00&  0.00&  0.04\cr
\qquad\alpha [2]   &-0.18&  0.03&  0.07&  0.01&  1.00& -0.02&  0.41&  0.01& -0.19&  0.01& -0.02&  0.01&  0.21\cr
\qquad\delta[2]    & 0.07& -0.01& -0.03& -0.00& -0.02&  1.00&  0.06&  0.96&  0.07&  0.00&  0.01&  0.00& -0.07\cr
\mu_\alpha\cos(\delta)[2]& 0.70& -0.10& -0.28& -0.03&  0.41&  0.06&  1.00& -0.04&  0.72&  -0.05&  0.07& -0.05& -0.8\cr
\qquad\mu_\delta [2] &-0.05&  0.01&  0.02&  0.00&  0.01&  0.96& -0.04&  1.00& -0.05& 0.00&  0.00&  0.00&  0.06]\cr
\qquad\alpha [3] & 0.79& -0.11& -0.32& -0.04& -0.19&  0.07&  0.72& -0.05&  1.00& -0.06& -0.34& -0.06& -0.90\cr
\qquad\delta[3]  &-0.05&  0.01&  0.02&  0.00&  0.01&  0.00& -0.05&  0.00& -0.06& 1.00& -0.01& -0.94&  0.06\cr
\mu_\alpha\cos(\delta)[3]& 0.07& -0.01& -0.03&  0.00& -0.02&  0.01&  0.07&  0.00& -0.34& -0.01&  1.00& -0.01& -0.08\cr
\qquad\mu_\delta [3] &-0.06&  0.01&  0.02&  0.00&  0.01&  0.00& -0.05&  0.00& -0.06& -0.94& -0.01&  1.00&  0.06\cr
\qquad\pi     &-0.88&  0.12&  0.35&  0.04&  0.21& -0.07& -0.80&  0.06& -0.90& 0.06& -0.08&  0.06&  1.00\cr
}
\] 
\end{small}

\subsection{With offset}
\[
\bordermatrix{
&\alpha [1]  & \mu_\alpha\cos(\delta) [1] & \mu_\alpha\cos(\delta) [2] & {\rm Offset}(\alpha) & \pi\cr
\qquad\alpha [1]         & 1.00 & -0.90 & -0.09 & -0.93 & -0.98 \cr
\mu_\alpha\cos(\delta)[1]&-0.90 &  1.00 &  0.08 &  0.82 &-0.83 \cr
\mu_\alpha\cos(\delta)[2]&-0.09 &  0.08 &  1.00 & -0.26 & -0.09 \cr
{\rm Offset}(\alpha)      &-0.93 & -0.82 &  0.26 &  1.00 & -0.93 \cr
\qquad\pi             &-0.98 & -0.83 & -0.09 & -0.93 &  1.00  \cr
}
\] 

\[
\bordermatrix{
&\delta [1]  & \mu_\delta [1] & \mu_\delta [2] & {\rm Offset}(\delta) & \pi\cr
\quad\delta [1]    & 1.00 & -0.75 & -0.78 & -0.84 & -0.98 \cr
\quad\mu_\delta[1] &-0.75 &  1.00 &  0.52 &  0.63 &-0.65 \cr
\quad\mu_\delta[2] &-0.78 &  0.52 &  1.00 &  0.33 &  0.80 \cr
{\rm Offset}(\delta)&-0.84 &  0.63 &  0.33 &  1.00 & 0.82 \cr
\quad\pi           &-0.98 & -0.65 &  0.80 &  0.82 &  1.00  \cr
}
\] 

\subsection{Spot 1 \& 2}

\[
\bordermatrix{
&\alpha & \delta & \mu_\alpha\cos(\delta) &\mu_\delta & \pi\cr
\qquad\alpha& 1.00 & -0.01 & -0.27 &  0.07 &  0.22 \cr
\qquad\delta&-0.01 &  1.00 &  0.04 & -0.21 & -0.04 \cr
\mu_\alpha\cos(\delta)&-0.27 &  0.04 &  1.00 & -0.30 & -0.96 \cr
\qquad\mu_\delta& 0.07 & -0.21 & -0.30 &  1.00 &  0.31 \cr
\qquad\pi & 0.22 & -0.04 & -0.96 &  0.31 &  1.00  \cr
}
\] \\

\subsection{Shock front}

\[
\bordermatrix{
&\alpha &  \mu_\alpha\cos(\delta) & \pi\cr
\qquad\alpha& 1.00 &  0.56 & -0.60 \cr
\mu_\alpha\cos(\delta)& 0.56 &  1.00 & -0.94 \cr
\qquad\pi &-0.60 & -0.94 &  1.00  \cr
}
\] \\
}

\section{Monte Carlo Simulations}\label{sec:MC}

To test the reliability of our results, particularly in relation to the short 
observing timespan, we performed the following simulations. 
We created new data sets consisting of simulated observed positions at the epochs
when the spots 1, 2, and 3 were detected. We then evaluated their positions following
the equations appropriated for a particle moving with a linear proper motion 
($\mu$) plus the trigonometric parallax ($\pi$), to be:

\begin{equation}
\begin{split}
P(0)=&X(0)+f_{\pi}(0)*\pi\\ 
P(1)=&X(0)+\mu*t(1)+f_{\pi}(1)*\pi\\ 
.&\\
.&\\
.&\\
P(n)=&X(0)+\mu*t(n)+f_{\pi}(n)*\pi 
\end{split}
\end{equation}

\noindent where the parameter are as follows. $P(n)$ is the simulated position at 
the epoch $n$. The start position, $X(0)$, is the position of our detected water
maser spots in their first detection. Finally, $f_{\pi}(n)$ is the projection of the 
parallactic ellipse at epoch $n$ \cite[e.g.,][]{urban2012}.

To each simulated position we added a random error drawn from a normal 
distribution of mean equal to zero and standard deviations of 0.4 mas and 0.8 mas 
in R.A. and Dec., respectively. These last values are similar to the maximum 
values of our systematic errors ($E_{\rm sys}$; see Table~\ref{tab:ffit}).
Each data set was then fitted using the same scheme we used for the actual 
observations. 

\begin{figure}[!h]
   \centering
  \hspace{-0.7cm}\includegraphics[height=0.43\textwidth, trim=30 0 0 0, clip]{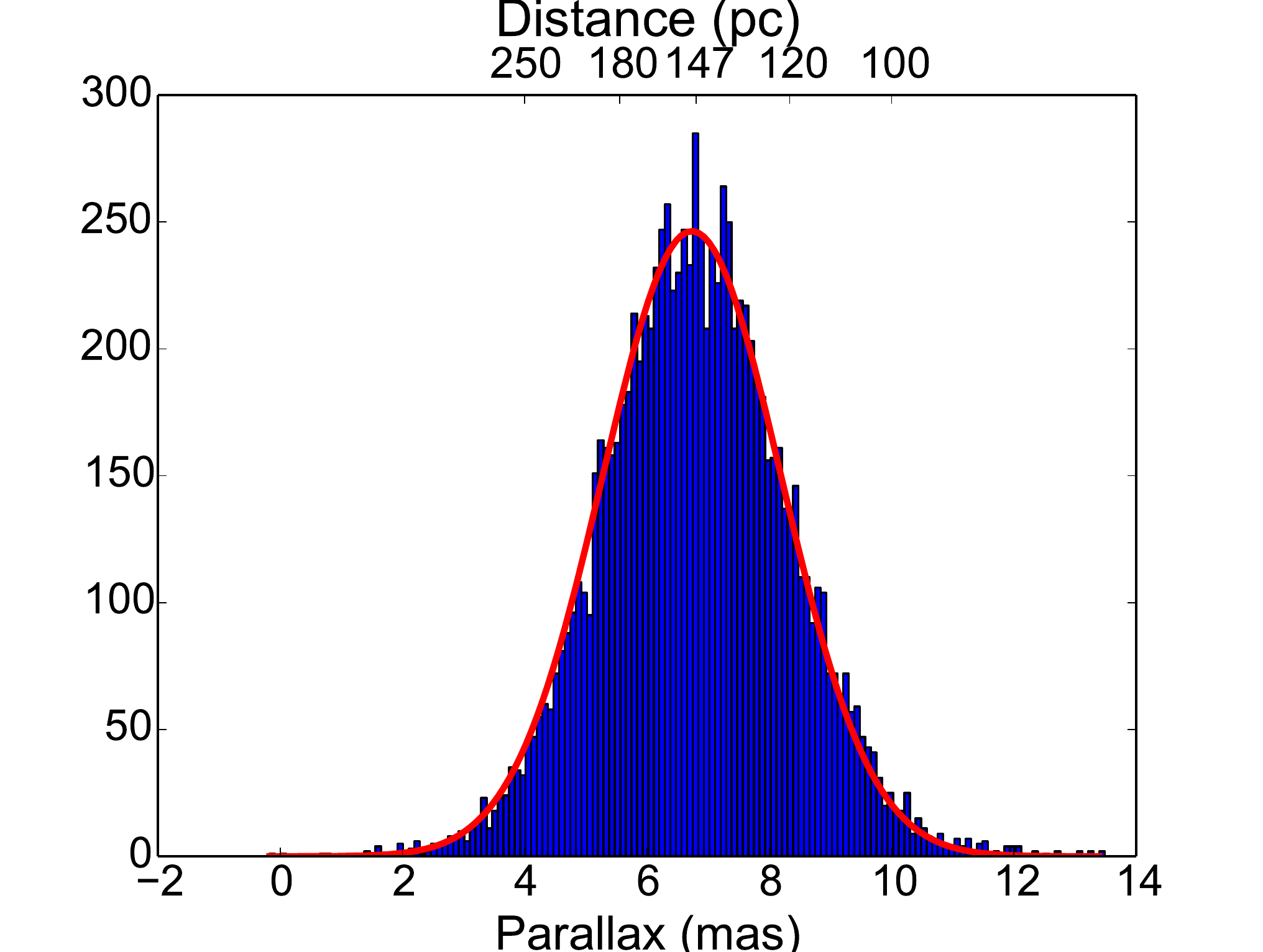} 
   \caption{Distribution histogram of the simulation set corresponding to the Simultaneous 1 set, at the input distance 147.3~pc.
}
   \label{fig:hist}
\end{figure}

We created 10000 realizations of the simulated positions and then calculated the 
mean and standard deviation  of the resulting distributions of the astrometric 
parameters. The input values used in the simulation as well as the resulting 
values are given in Table~\ref{tab:MCffit}. As an example, a distribution histogram 
of one of the simulation sets is presented in Figure~\ref{fig:hist}.

\begin{table*}[!ht]
\small
\renewcommand{\arraystretch}{1.31}
  \begin{center}
  \caption{Input and output parameters of the Monte Carlo simulations.  }
    \begin{tabular}{cc|cccc|cc|cc|cc|cc}\hline\hline
     &            &  \multicolumn{4}{c|}{Input parameters}&  \multicolumn{8}{c}{Output parameters}\\
     &            &  $\pi$   &  d  & $\mu_{\alpha}\cos{\delta}$ &  $\mu_{\delta}$  &  \multicolumn{2}{c}{$\pi$}   &  \multicolumn{2}{c}{d}  & \multicolumn{2}{c}{$\mu_{\alpha}\cos{\delta}$} &  \multicolumn{2}{c}{$\mu_{\delta}$}    \\
  &  Test      &  (mas)   &(pc) & (mas yr$^{-1}$) & (mas yr$^{-1}$)   &  \multicolumn{2}{c}{(mas)}   & \multicolumn{2}{c}{(pc) } & \multicolumn{2}{c}{(mas yr$^{-1}$)} & \multicolumn{2}{c}{(mas yr$^{-1}$)}     \\
Fit  & particle       &     &    &   &     &  Mean & SD &Mean  & SD & Mean & SD & Mean & SD     \\
\hline
\multirow{9}{*}{Individual}& 1 & \multirow{3}{*}{8.3333}   &\multirow{3}{*}{120}&\multirow{3}{*}{$-40$}&\multirow{3}{*}{$-20$}& 8.1 & 3.7 &124 &57 & $-40.2$ & 3.6  & $-20.5$ & 7.4\\
 & 2  &   &   &  &  & 8.3 & 1.5 & 121 & 22 & $-40.1$ & 2.2  & $-20.2$ & 4.5\\
 & 3  &   &   &  &  & 8.2 & 2.3 & 122 & 34& $-40.3$ & 3.5  & $-20.3$ & 5.4\\ \cline{2-14}
 & 1 & \multirow{3}{*}{6.7889}   &\multirow{3}{*}{147.3}&\multirow{3}{*}{$-40$}&\multirow{3}{*}{$-20$}&6.8 & 3.9 &148 &85&$-40.1$ & 3.5  & $-20.3$ & 7.3\\
 & 2  &   &   &  &  &  6.7& 1.5 & 149 & 34 & $-40.1$ & 2.2  & $-20.2$ & 4.5\\
 & 3  &   &   &  &  &  6.8& 2.3 & 147 & 50& $-40.1$ & 3.6  & $-20.1$ & 5.6\\ \cline{2-14}
 & 1 & \multirow{3}{*}{5.5556}   &\multirow{3}{*}{180}&\multirow{3}{*}{$-40$}&\multirow{3}{*}{$-20$}&5.4 &3.8 &184 & 129&$-40.2$ & 3.5  & $-20.4$ & 7.5\\
 & 2  &   &   &  &  & 5.5 & 1.5 & 182 &51 & $-40.1$ & 2.2  & $-20.2$ & 4.7\\
 & 3  &   &   &  &  & 5.4 & 2.3 & 185 &78 &$-40.1$ & 3.5  & $-20.3$ & 5.7\\ 
\hline
\multirow{9}{*}{Simultaneous 1}& 1 & \multirow{3}{*}{8.3333}   &\multirow{3}{*}{120}&\multirow{3}{*}{$-40$}&\multirow{3}{*}{$-20$}
 &\multirow{3}{*}{8.3} & \multirow{3}{*}{1.5}  &\multirow{3}{*}{120} & \multirow{3}{*}{21} & $-40.3$ & 7.0  & $-20.5$ & 7.7\\
 & 2  &   &   &  &  &  & &  &  & $-40.1$ & 2.2  & $-20.3$ & 4.4\\
 & 3  &   &   &  &  &  & &  &  & $-40.1$ & 3.1  & $-20.3$ & 5.6\\ \cline{2-14}
 & 1 & \multirow{3}{*}{6.7889}   &\multirow{3}{*}{147.3}&\multirow{3}{*}{$-40$}&\multirow{3}{*}{$-20$}
  &\multirow{3}{*}{6.7} & \multirow{3}{*}{1.5}  &\multirow{3}{*}{148} & \multirow{3}{*}{32} & $-40.2$ & 6.8  & $-20.2$ & 7.8\\
 & 2  &   &   &  &  &  &  & & & $-40.0$ & 2.2  & $-20.2$ & 4.5\\
 & 3  &   &   &  &  &  &  & & & $-40.1$ & 3.1  & $-20.2$ & 5.6\\ \cline{2-14}
  & 1 & \multirow{3}{*}{5.5556}   &\multirow{3}{*}{180}&\multirow{3}{*}{$-40$}&\multirow{3}{*}{$-20$}
   &\multirow{3}{*}{5.5} & \multirow{3}{*}{1.5}  &\multirow{3}{*}{180} & \multirow{3}{*}{48} & $-40.3$ & 6.8  & $-20.3$ & 7.6\\
 & 2  &   &   &  &  &  &  & & & $-40.1$ & 2.2  & $-20.2$ & 4.6\\
 & 3  &   &   &  &  &  &  & & &$-40.1$ & 3.1  & $-20.4$ & 5.5\\ 
 \hline
 
\multirow{9}{*}{Simultaneous 2}& 1 & \multirow{3}{*}{8.3333}   &\multirow{3}{*}{120}&$-45$&$-25$
 &\multirow{3}{*}{8.3} & \multirow{3}{*}{1.4}  &\multirow{3}{*}{120} & \multirow{3}{*}{21} & $-45.6$ & 6.8  & $-25.0$ & 8.0\\
 & 2  &   &   & $-35$ & $-15$  &  & &  &  & $-35.1$ & 2.2  & $-15.1$ & 4.6\\
 & 3  &   &   & $-40$ & $-20$  &  & &  &  & $-40.1$ & 3.2  & $-20.1$ & 5.5\\ \cline{2-14}
 & 1 & \multirow{3}{*}{6.7889}   &\multirow{3}{*}{147.3}&$-45$&$-25$
  &\multirow{3}{*}{6.8} & \multirow{3}{*}{1.5}  &\multirow{3}{*}{147} & \multirow{3}{*}{32} & $-45.5$ & 7.0  & $-25.3$ & 7.8\\
 & 2  &   &   & $-35$ & $-15$ &  &  & & & $-35.1$ & 2.2  & $-15.2$ & 4.5\\
 & 3  &   &   & $-40$ & $-20$ &  &  & & & $-40.1$ & 3.0  & $-20.1$ & 5.6\\ \cline{2-14}
  & 1 & \multirow{3}{*}{5.5556}   &\multirow{3}{*}{180}&$-45$&$-25$
   &\multirow{3}{*}{5.5} & \multirow{3}{*}{1.5}  &\multirow{3}{*}{180} & \multirow{3}{*}{57} & $-40.2$ & 3.6  & $-20.5$ & 7.4\\
 & 2  &   &   & $-35$ & $-15$ &  &  & & & $-40.1$ & 2.2  & $-20.2$ & 4.7\\
 & 3  &   &   & $-40$ & $-20$ &  &  & & &$-40.1$ & 3.5  & $-20.3$ & 5.7\\ 
 \hline\hline
    \label{tab:MCffit}
    \end{tabular}
  \end{center}
\end{table*}

\end{appendix}

\end{document}